\documentclass[11pt,a4]{article}
\usepackage[applemac]{inputenc}  
\usepackage{amssymb,amsmath,amsthm,amsfonts,amscd,amstext}
\usepackage[mathscr]{eucal}
\usepackage{array}
\usepackage{graphicx}
\newtheorem{lem}{Lemma}[section]

\newtheorem{thm}[lem]{Theorem}

\newtheorem{df}{Definition}[section]

\newcommand{\Aut}{\mathrm{Aut}}
\def\aut{\operatorname {Aut}} 
\def\C{{\mathbb C}}  

\def\D{{\mathbb D}}

\def\der{\operatorname {Der}} 

\def\Diff{\operatorname {Diff}}

\newcommand{\e}{{\mathrm e}} 

\def\HH {\mathbb H} 

\def\OO{\cal O}  

\def\R{{\mathbb R}}    

\def\vir{\operatorname{Vir}} 

\def\Z{{\mathbb Z}}    
\def\N{{\mathbb N}}    

\title{\bf A Renormalisation Group approach to Stochastic L{\oe}wner Evolutions and the Doob $h$-transform}
\author
{Roland Friedrich\\
\\
\normalsize{MPI}\\
\normalsize{53111 Bonn, Germany}\\
\normalsize{e-mail: rolandf@mpim-bonn.mpg.de}
}
\begin{document}
\maketitle
\begin{abstract}
In this notes we shall describe the relation of a certain class of simple random curves 
arising in
2D statistical mechanics models in the scaling limit, which can be described dynamically 
by
Stochastic L{\oe}wner Evolutions (SLE), and the equivalent Renormalisation Group (RG) 
theoretic
interpretation in Conformal Field Theory, as a fixed point of the RG flow. Further, we 
shall recall the relation of this random curves with String Theory, and how one can 
derive a general measure on such random paths, by using weighted regularised 
determinants, which come from sections of twisted line bundles. Importantly, the null vector at level two in the Verma module for the highest-weight representation of the Virasoro algebra corresponds to a generalised Doob-Getoor $h$-transform.
\end{abstract}
\tableofcontents
\section{Introduction}
This text is a somewhat extended version of a talk I gave at the workshop ``RENORMALIZATION” at the Max-Planck-Institute for Mathematics in
Bonn, in December 2006.  

Its aim is to explain to a non-specialist audience, how Stochastic L{\oe}wner Evolutions (SLE) are connected with the Renormalisation Group (RG), but also what the underlying global geometric picture is. In doing so, I shall restrict myself to the case of the unit disc, the original model, and leave out the general theory for arbitrary Riemann surfaces, as developed in~\cite{F, FK, K, KS}. During the process of writing, I felt free to expand on results which I obtained several years ago. 

There has been two major directions to construct measures on random loops or intervalls, namely the one pursued by P.~Malliavin~\cite{M} and the other one by O. Schramm, who initiated SLE~\cite{Sch}, but also G. Lawler, W. Werner~\cite{LSW}, (LSW), and M. Aizenman~\cite{AB}. Both was taking place roughly around 1999.  
The (technical) foundations of SLE were laid down by Schramm and S.~Rohde in~\cite{RSch}, and subsequently applied by LSW to open probabilistic problems.

The parts of the theory underlying SLE which we are going to present here, aims at giving a unified treatment which is capable to incorporate both approaches mentioned at the beginning, but also to explain all the general results physically, as proposed e.g., in~\cite{LSW}. The main tool to achieve this are determinant line bundles, i.e. domain-wise regularised determinants, and the so-called ``Virasoro Uniformisation" (VU), which had been co-invented by M.~Kontsevich, quite some time ago, and which generalises parts of the work of Kirillov, Yurev and Neretin. Also, (VU) is at the heart of the fundamental approach to CFT on Riemann surfaces, by Tsuchiya, Ueno and Yamada~\cite{TUY}.

Finally, from a mathematical perspective the main point is that the null vector at level two in the Verma module of the highest-weight representation of the Virasoro algebra corresponds to the Doob-Getoor $h$-transform. To consider harmonic sections with respect to the generator of the Diffusion process on the determinant line bundle is required if one wants to have martingales. In turn martingales are dictated by the physical set-up, namely to derive from models of Statistical Mechanics in a dynamical fashion measures on simple paths. 

A detailed and comprehensive companion text to the material presented here, is~\cite{F1}.

\section{Different point of views}
\begin{figure}[htbp!]
\begin{center}
\includegraphics[scale=0.25]{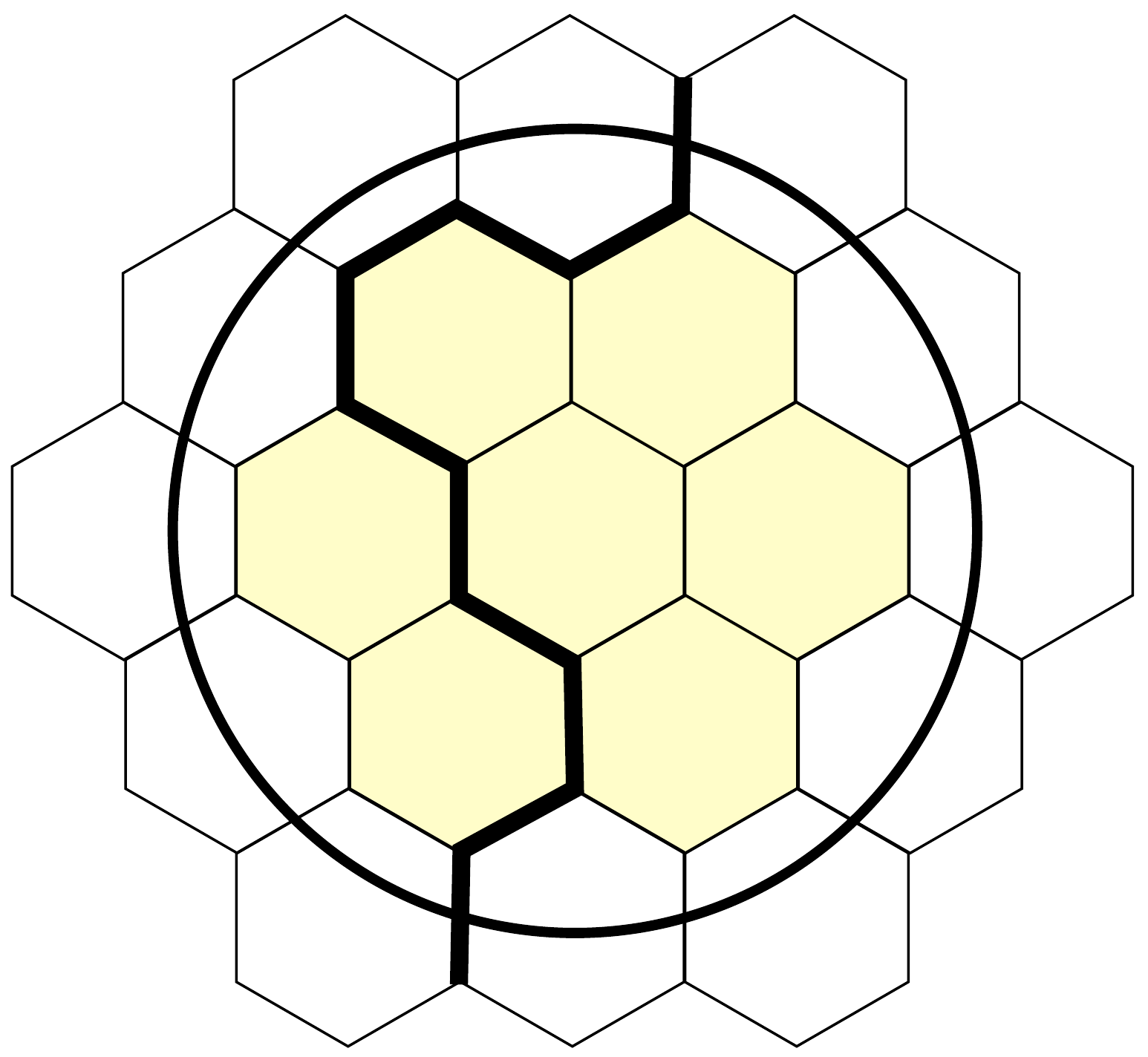}
\caption{A simple curve on a hexagonal lattice connecting two boundary points. Topological set-up.} 
\label{SAW}
\end{center}
\end{figure}

Let us start with the following example, as depicted in Figure~\ref{SAW}. There we have a hexagonal lattice approximating a closed disc. The yellow coloured hexagons compose the interior in the discretisation of the  disc and the white ones, the boundary, indicated by the circle.  
Further we have drawn a simple polygonal path connecting two boundary points and  strictly laying in the interior of the disc. 

From a discrete, combinatorial and enumerative point of view, we can ask several natural questions. E.g. how many discrete simple path are connecting the same two boundary points, thereby always running inside the circle or what is the (Euclidean) length of the longest path. It is important to note, that here boundary and interior are a priori just  topological notions, merely giving the roughest binary framework to start with. 

If we would keep the circle fixed but replace our lattice by another of a finer grid size, we still can ask the same questions but some of them become increasingly unnatural. Also, the combinatorial complexity of the problem would start to explode, so that instead of keeping track of all possibilities, we would be forced to resort to statistical methods. But what does that exactly mean and what are the relevant quantities we should look for.

Well, by now many people and groups invented their questions and methods, as the above problem could arise e.g. in Polymer  Physics. 

A particular challenge would be to construct probability measures on such discrete paths and to understand how all this scales as the mesh size becomes smaller and smaller. Also, what should the continuum object be and what are the natural measures to start with, as the same set can support different ones.

Again, depending on the context, e.g. dynamical generation of paths or ``static sets", several classes of measures have been considered and proposed. 

\begin{figure}[htbp!]
\begin{center}
\includegraphics[scale=0.25]{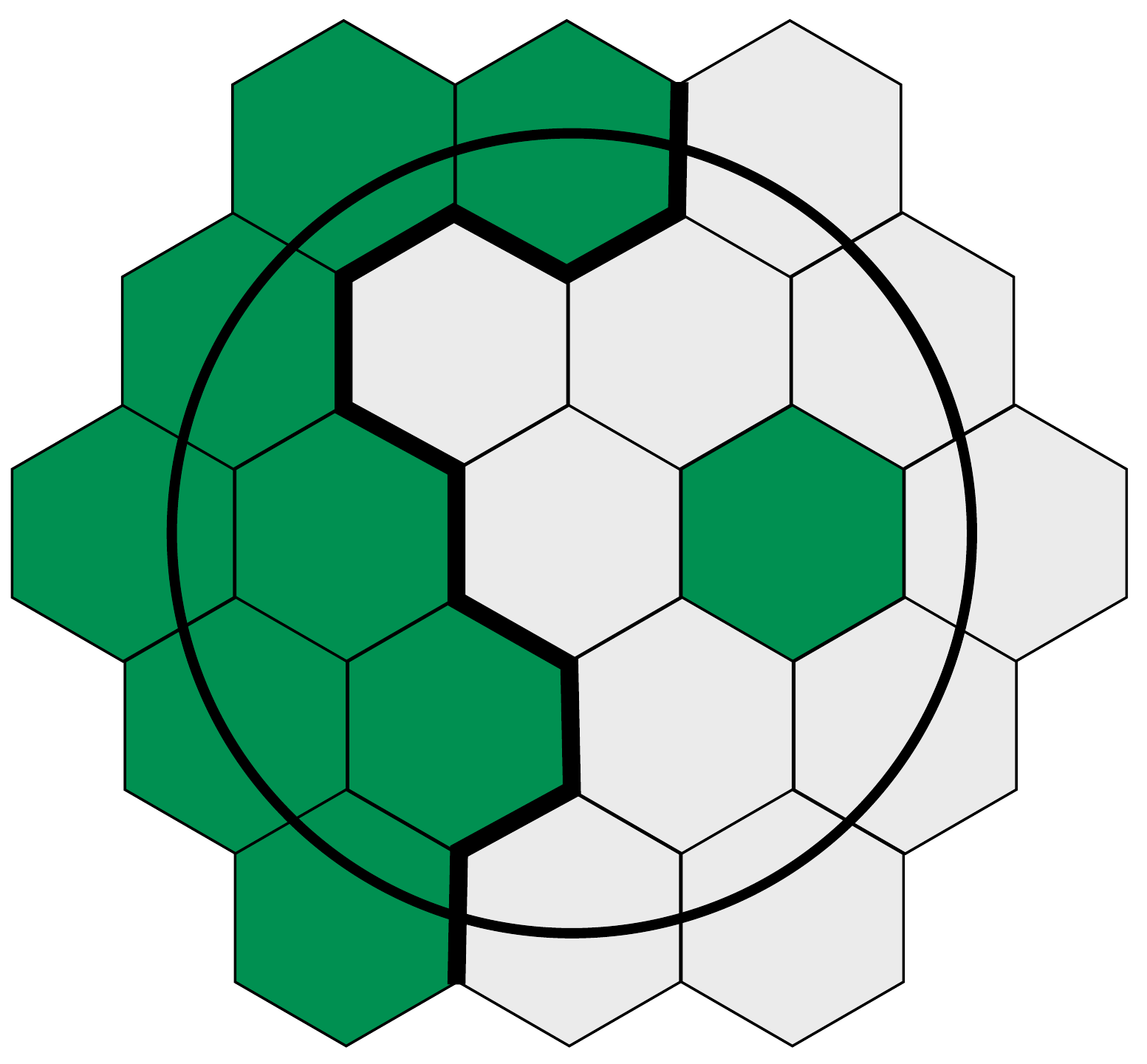}
\caption{A simple curve on a hexagonal lattice connecting two boundary points. Boundary condition set-up.} 
\label{SAW_Boundary}
\end{center}
\end{figure}

As we shall discuss in the next section, Statistical Mechanics and in particular Lattice Models, provide a rich source of such curves in form of interfaces. But, there are some fine points one has to be aware of, as the simple topological set-up will be altered by the notion of boundary conditions, as depicted in Figure~\ref{SAW_Boundary}, and further explained latter. However, boundary conditions intrinsically reveal that we are taking up a macroscopic stance on the problem, which shifts our focus on what to consider as being natural and interesting. 

\section{Motivation from Statistical Mechanics}
Besides percolation, as originally, the two-dimensional Ising model has by now become a widely adapted way in the literature to motivate SLE and to connect it with statistical mechanics respectively CFT,~\cite{FK, F, K}. 

So let us start with a simply connected and bounded domain $D$ in the plane $\mathbb{R}\cong\mathbb{C}$ and a triangular lattice (TL) of mesh size $\delta>0$, ``just" covering the closed set $\overline{D}$ . We shall assume that the boundary of the domain is sufficiently smooth and the  mesh size enough fine to avoid cumbersome complications.

A configuration of spins is a function $\sigma$ on the set of vertices $V(TL)$ with values $\pm 1$. We may now pass to the dual hexagonal lattice, and colour the hexagons enclosing a vertex with value $-1$ black and those with value $+1$ white. In the percolation setting this would correspond to site (or vertex) percolation on a triangular grid.

\begin{figure}[htbp!]
\begin{center}
\includegraphics[scale=0.4]{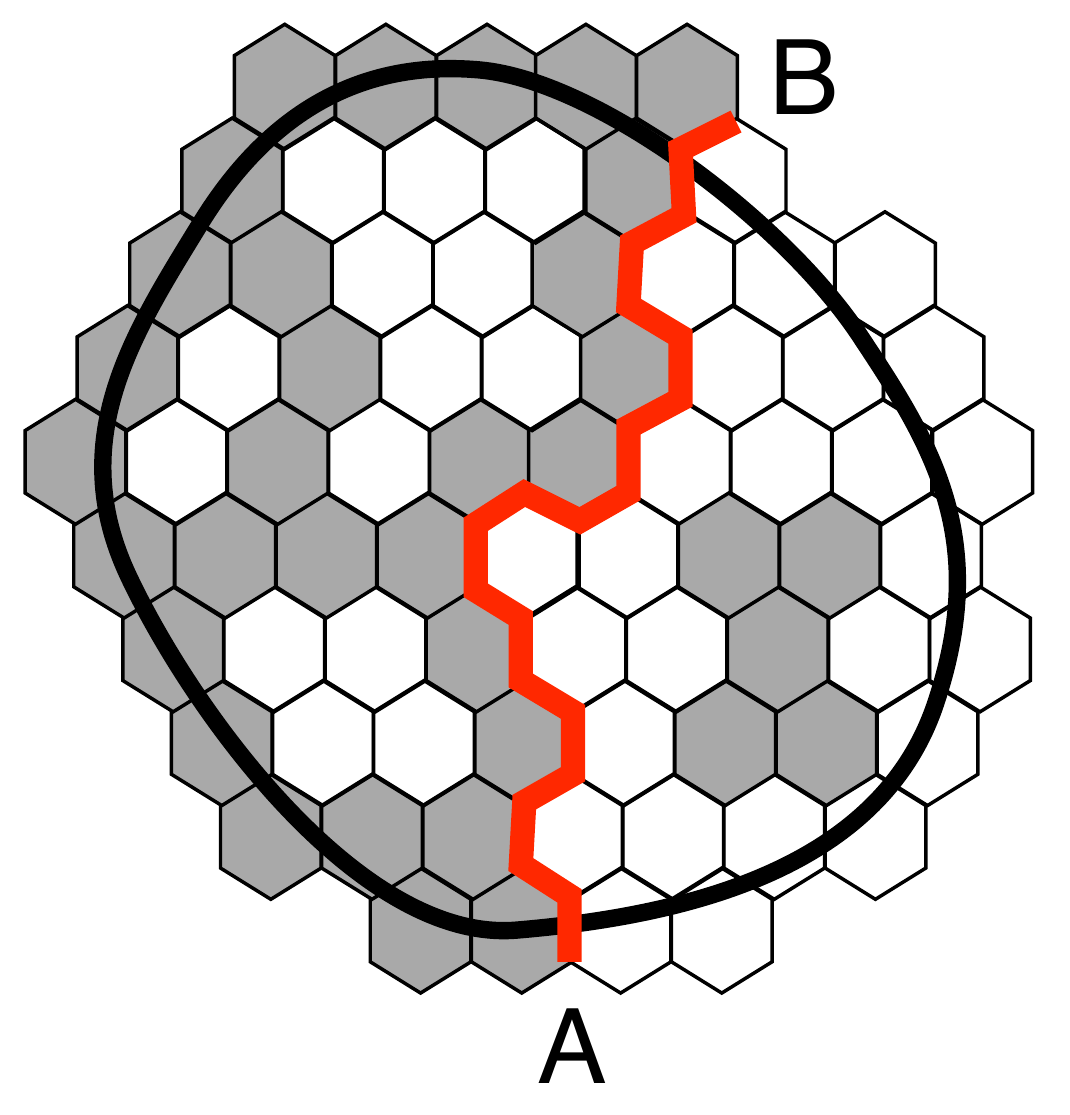}
\caption{The interface associated with the two-dimensional Ising model.} 
\label{Hexagon}
\end{center}
\end{figure}

A sure way to get an interface (domain wall), as illustrated in Figure~\ref{Hexagon}, is to take two marked points $A$ and $B$ on the  boundary $\partial D$ and then to fix the value of the spin sitting in the centre of exactly one hexagon black, if that one intersects the boundary segment from $B$ to $A$ and similarly white, for that boundary part which belongs to the segment from $A$ to $B$, assuming a counter-clockwise orientation.

The curve which arises is a Jordan  curve (no branching) connecting vertices such that on the left we always have spin $-1$ and to the right $+1$. 

In the case of our model with this particular choice of boundary conditions, we  have an object that  persists on all finite scales  and should therefore  be a good candidate to proceed to an object which can be regarded as a ``natural" scaling limit, i.e. a macroscopic observable. 

However as a look at Figure~\ref{Hexagon} reveals, once the particular (red) interface is fixed, we could change some of the values of the spins corresponding to boundary parts, without altering the shape of the curve itself, which implies that the set of all configurations having that particular interface, is much bigger than just those compatible with the fixed ``domain wall" boundary conditions.

\begin{figure}[htbp!]
\begin{center}
\includegraphics[scale=0.4]{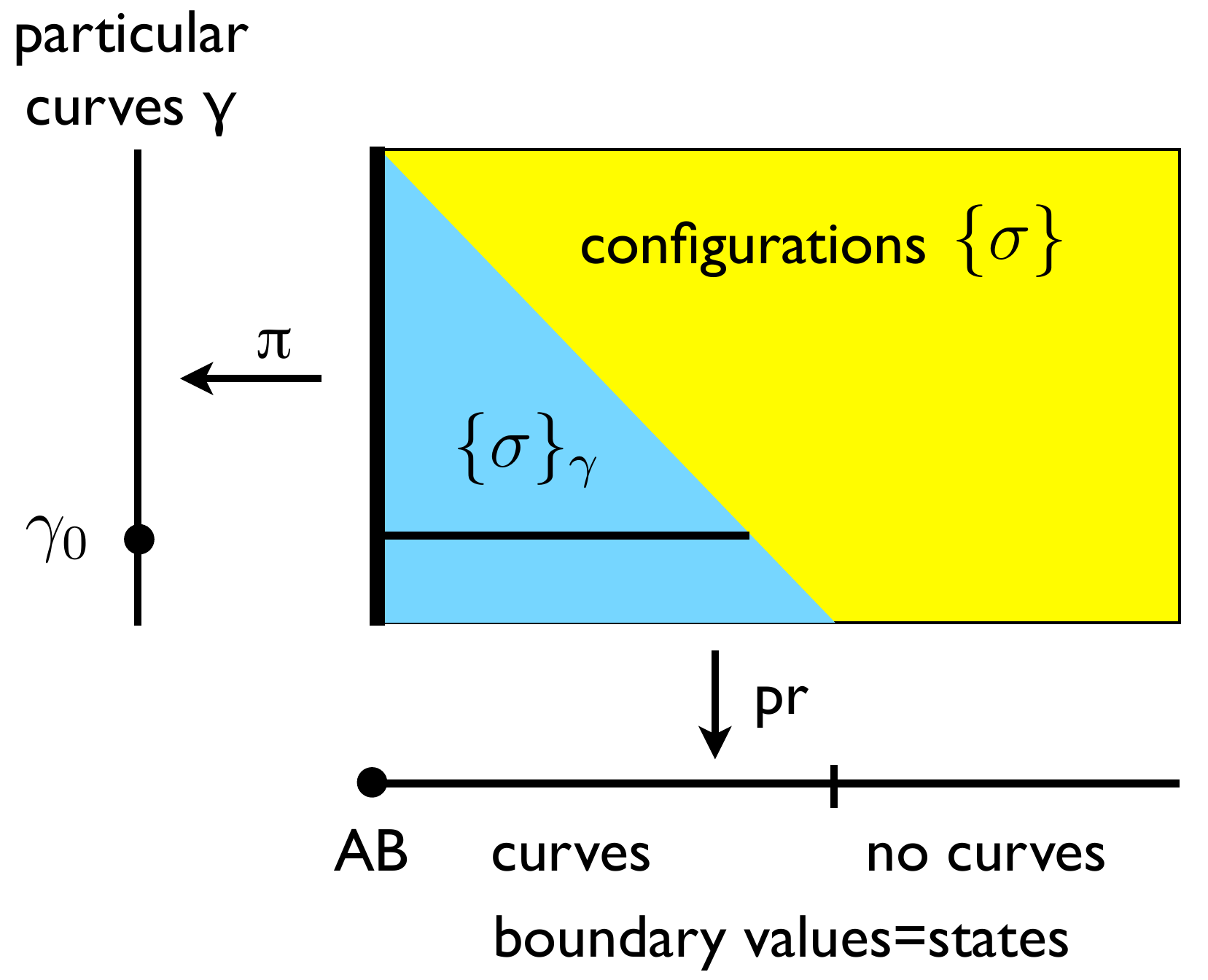}
\caption{The configuration space as a fibered space over the state space, principal fibration, or the non-local fibration over curves.} 
\label{configurations}
\end{center}
\end{figure}

The situation at the discrete level is summarised schematically in Figure~\ref{configurations}.  The (yellow-blue) square represents the space of all possible configurations, which is fibered over the space of all possible boundary values, i.e. the state space, and corresponds to the case of ``free boundary values". 

The state space can be subdivided into two disjoint non-empty sets, which are labelled as ``curves" respectively  ``no curves" and which contain those boundary conditions which may produce a curve $\gamma$ connecting the two marked points $A$ and $B$, respectively not. It is important to note that the fibre over a state from ``curves" may contain configurations which do not include a domain wall running from $A$ to $B$. This is indicated by the (blue) triangle $\{\sigma\}_{\gamma}$ over the set ``curves". 

There is one special state, labelled $AB$, which corresponds to the boundary conditions described as ``domain wall boundary conditions", (cf. Fig.~\ref{Hexagon}). Now, the fibre over the state $AB$ is composed of configurations that do always contain a Jordan arc running from $A$ to $B$ and therefore we may call it the ``special fibre".

We have yet another (partial) fibration. Namely the space of configurations with a domain wall (blue triangle) is fibered  over the set of particular curves. In our Figure~\ref{configurations}, this is depicted by the vertical line labelled ``particular curves" with the projection map $\pi$. It is important to note, that say for a curve $\gamma_0$, the intersection of the fibres $\pi^{-1}(\gamma_0)$ and $\text{pr}^{-1}(AB)$ is not just one point (configuration), but indeed does contain several.
\subsection{Gibbs measures with ``proper" boundary conditions}
The Gibbs measure on configurations naturally induces a probability measure on the random curves we just considered. However, the boundary conditions we choose, play  an essential role.

So, in the thermodynamical equilibrium at the absolute  temperature $T>0$ the Gibbs measure is described by  
\begin{displaymath}
d\mu_G(\sigma):=\frac{1}{Z} e^{-\beta E[\sigma]},\qquad \beta:=\frac{1}{T}\quad\mbox{(inverse temperature)}~,
\end{displaymath} 
where the functional $E[\sigma]$, denotes the energy of a configuration and the numerical pre-factor $Z>0$  serves as normalisation, such that the sum of all elementary events ads up to 1. It is defined as 
\begin{displaymath}
Z:=\sum_{\{\sigma\}} e^{-\beta E[\sigma]}~,
\end{displaymath}
with the sum extending over all configurations $\{\sigma\}$ and it is called the partition function. The measure for a fixed domain $D$ depends on two parameters, namely the temperature but also the underlying mesh size, i.e. $\mu_G=\mu_{G(T,\delta)}$.

If we choose fixed $AB$-boundary conditions for the calculation of the partition function, then we are restricting ourselves to the special fibre over $AB$, and take its Boltzmann-weighted volume $Z_{{\sigma}_{AB}}$ as normalisation factor. For free boundary conditions we would restrict the summation to the set of all configurations which contain a domain wall, i.e. to the set $\{\sigma\}_{\gamma}$ (blue triangle) with $Z_{{\sigma}_{\gamma}}$ as the corresponding partition function. 

The natural equivalence relation on the set of configurations $\{\sigma\}_{\gamma}$  is given by declaring two realisations $\sigma_1$ and $\sigma_2$ as being equivalent if they include the same chordal domain wall, connecting the  two marked  boundary points, cf. Fig.~\ref{Hexagon}. Again, we can restrict the equivalence relation to the fibre $\text{pr}^{-1}(AB)$ only. In any case, the situation yields the fibration with base space the particular curves and the equivalent configuration corresponding to the fibres, respectively to their intersection with $\text{pr}^{-1}(AB)$. 

The probability measure on the quotient space, with simple events the particular curves $\gamma$, is the image measure, either with respect to $Z_{{\sigma}_{AB}}$ or $Z_{{\sigma}_{\gamma}}$, i.e. $\pi_*\mu_{G_{AB}}$ resp. $\pi_*\mu_{G_{\sigma_{\gamma}}}$.

Although the geometric set of simple random curves is the same (topological set-up), by taking different boundary conditions we arrive at having two different measures on the same set.

Now, and this is important to note, the proofs (analytic or numeric for the Ising model)~\cite{Sm, GC} concerning the scaling limit of the above measures, i.e. the approach of the thermodynamic limit along the critical temperature $T_c$,  deal with the family of measures on the special fibre 
$\text{pr}^{-1}(AB)$. 

Further, as we shall discuss in the next section, on general grounds it was / is expected that the scaling limit is a measure, supported on simple paths, which is conformally invariant. 

Let us close with the following remarks. The procedure just described, is not particular to the Ising model.  We can choose other  models, e.g. $Q$-states Potts model (i.e. different Boltzmann-weights) as well as other lattices (not necessarily hexagonal) with appropriate boundary conditions (e.g. wired and free). For lattice models there are  other  ``good" probabilistic events. In the Ising model slightly below the critical temperature $T_c$ we observe the sea of nested Jordan curves of domain boundaries.
\begin{figure}[htbp!]
\begin{center}
\includegraphics[scale=0.5]{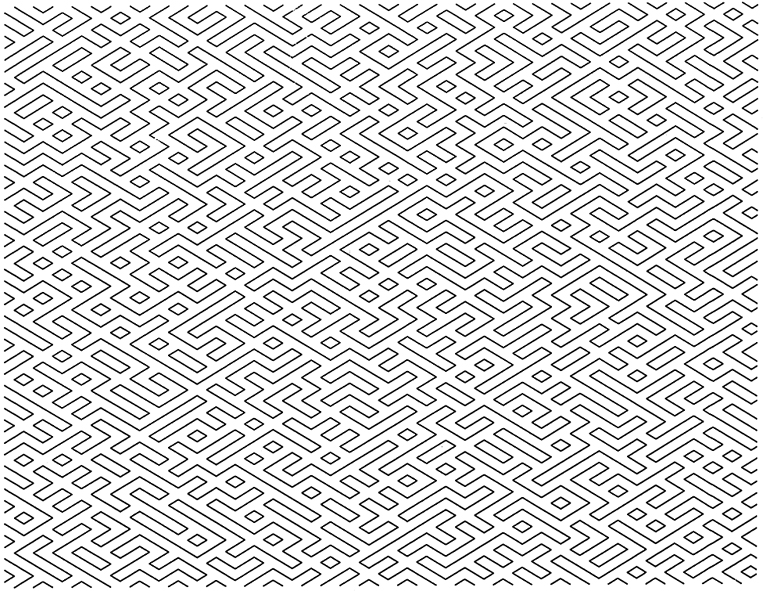}
\caption{Central charge zero random field of loops on a torus, which has a dual description as a percolation model. (Result from unpublished work on Random Fields and SLE, 2000/2001)} 
\label{cd}
\end{center}
\end{figure}
Again,  by passing to $T_c$ and simultaneously rescaling we obtain a dense collection of closed, non-intersecting loops, carrying a scale invariant probability distribution which would be also derived from the series of Gibbs measures as we explained in the case of chordal lines. (cf.~\cite{KS})

\section{Measures as Domain Functionals, the ALPS-correspondence and SLE}
So-far our discussion centred around a fixed domain and the measures associated with it, supported on simple curves and connecting two previously chosen distinct boundary points. Now, we could ask, how the measure would change if we keep the domain, but vary the points, or even more generally, also vary the underlying domains. So, what we have in fact is a domain functional
\begin{equation}
\label{dom_funct}
D_{A,B}\mapsto\mu_{D_{A,B}}
\end{equation}
where $D$ denotes the domain, and $A,B\in\partial D$, two distinguished (and ordered) boundary points.

In two dimensions, there exists a very strong statement, the Riemann mapping theorem, which tells that any two simply connected domains are conformally related, and if the domains are bounded by Jordan curves, then this mapping can be extended as a homeomorphism to the whole boundary, according to Carathéodory's theorem. 

Now it was conjectured~\cite{Ai, LPS}, that these measures should be conformally invariant in the scaling limit, which can be expressed as the following commutative diagram, 
\[
\begin{CD}
D_{A,B}@>f>> D'_{A',B'}\\
@V F_{\text{ALPS}} VV  @V F_{\text{ALPS}}VV\\
\mu_{D_{A,B}} @>f_*>> \mu_{D'_{A',B'}}
\end{CD}
\]
with the morphism $f$ being a conformal equivalence which respects the ordered marked points, i.e. $f(A)=A'$ resp. $f(B)=B'$, and as objects Jordan domains with two marked and ordered points, $\text{\texttt{JDom}}_{\bullet,\bullet}$~. As we shall see latter, this category can be enlarged to include also slit domains. $f_*$ denotes the induced measurable mapping, i.e. $F_{\text{ALPS}}(f)=:f_*$, where $F_{\text{ALPS}}$ stands for the ``Aizenman--Langlands--Pouilot--Saint-Aubin" correspondence (functor).

Let us note that because of conformal invariance and because the set of morphisms between any two objects in $\text{\texttt{JDom}}_{\bullet,\bullet}$ has more than one object, in fact a real continuum, it is enough to study just one reference model,  e.g. the upper half-plane $\mathbb{H}$ with $0$ and $\infty$ as the marked points.

The category $\text{\texttt{JDom}}_{\bullet,\bullet}$ has several equivalent parametrisations. Namely, if we consider first the set 
$\text{\texttt{J}}^{\infty}$ of smooth Jordan curves,  then one has the double quotient~\cite{AMT}
$$
\text{SU}(1,1){\setminus}\Diff_+(S^1) / \text{SU}(1, 1)
$$
as the base space, and as fibre model of the total space, the torus  (minus the diagonal), i.e., $S^1\times S^1\setminus\{\text{diagonal}\}$.
Sections then correspond to domains with two distinct marked boundary points.

As these measures are translational invariant, it is enough to look at $\text{\texttt{J}}^{\infty}_0$, i.e., all smooth Jordan curves surrounding the origin.  The uniformising application with domain the unit disc $\D$, is unique if we require it to preserve the origin $0$, and to have either a strictly positive derivative at $0$ or to map $1$ onto one of the marked points. 

For symmetry reasons we take the first normalisation. Then the  manifold $\cal M$ of all such univalent maps corresponds to a contractible subset in~(cf.~\cite{F})
$$
\Aut({\OO})=\{a_1z+a_2 z^2+a_3 z^3\dots,~ 0<a_1\}\hookrightarrow\C^{\N^{*}}, 
$$ 
which itself is a contractible space. By the Bieberbach-DeBranges theorem the coefficients of elements in $\cal M$ satisfy $a_n\leq n\cdot a_1$. Similarly, we have

\[
\begin{CD}
\{\text{conformally invariant measures}\}_{\kappa>0}@>>>{\cal M}\times (S^1\times S^1\setminus\{\text{diagonal}\})@>\pi>>{\cal M}\hookrightarrow\Aut({\OO})~.
\end{CD}
\]
Then the trivial bundle, with fibre (almost) a torus, parametrise, up to one real positive constant $\kappa$, conformally invariant probability measures on simple paths which connect two distinct boundary points, as we shall see. 

It was O.~Schramm's original insight~\cite{Sch}, to use L{\oe}wner's slit mapping to describe these random traces dynamically and to classify then all possible conformally invariant measures on them, i.e. to show the existence of $\kappa$. 

Physically, the existence of such a parameter can be understood as labelling models from Statistical Mechanics that contribute a measure, as previously explained. However, it is a deep and fundamental fact of the two-dimensional conformally invariant realm, that one parameter is enough, as we shall see in Section~\ref{RG_sec}.

\subsection{Schramm-L{\oe}wner Evolutions}
One of the basic observations in deriving the ``driving function" in the dynamical approach to random Jordan arcs is, besides a symmetry argument, continuity and the previously introduced  notion of conformal invariance, the following Markovian type property (stated for the category $\text{\texttt{JDom}}_{\bullet,\bullet}$):  

For a domain $D$ with non-degenerate boundary, let  ${\cal{W}}(D_{A,B})$ be the set of Jordan arcs in $D$ with endpoints $A$ and $B$. Denote by $\{\mu_{D_{A,B}}\}$ a family of probability measures on Jordan arcs in the complex plane such that 
\[
	\mu_{D_{A,B}}({\cal W}(D_{A,B})=1~.
\]
Then the Markovian-type property says for $\gamma$ a random Jordan arc,  that if $\gamma'$ is a sub-arc of $\gamma$ which has $A$ as one endpoint and whose other endpoint we denote by $A'$, then the conditional distribution of $\gamma$ given $\gamma'$ is 
\begin{equation}\label{E:markov}
\mu_{D_{A,B}|_{\gamma'}}=\mu_{(D\backslash\gamma')_{A',B}}~.
\end{equation} 

So, the only compatible driving function which satisfies the above requirements has to be proportional to standard one-dimensional Brownian motion, which leads to the following facts~\cite{Sch, RSch}.

The  chordal SLE$_\kappa$ curve $\gamma$ in the upper half-plane $\HH$ describes the growth of simple random curves emerging from the origin and aiming at infinity, as follows:
\begin{df}[Stochastic L{\oe}wner Equation]
\label{loewner_eq}
For $z\in\HH$, $t\geq0$ define $g_t(z)$ by $g_0(z)=z$ and 
\begin{equation}
\frac{\partial g_t(z)}{\partial t}  = \frac{2}{ g_t(z) - W_t}
\label{lowner}.
\end{equation}
\end{df}
The maps $g_t$ are normalised 
such that $g_t (z) = z + o(1) $ when $z \to \infty$
and  $W_t:= \sqrt{\kappa}\,B_t$ where  $B_t(\omega)$ is the standard
one-dimensional Brownian motion, starting at 0 and with variance
$\kappa>0$. Given the initial point $g_0(z)=z$, the ordinary differential
equation (\ref{lowner}) is well defined until a random time
$\tau_z$ when the right-hand side in (\ref{lowner}) has a pole.
There are two sets of points that are of interest, namely the preimage of infinity $\tau^{-1}(\infty)$ and its complement. For those in the complement we define:
\begin{equation}
\label{Khull}
K_{t}:=\overline{\{z\in\HH: \tau(z)<t\}}
\end{equation}
The family $(K_t)_{t\geq0}$, called  hulls, is an increasing family of compact sets in $\overline{\HH}$ where $g_t$
is the uniformising map from $\HH\setminus K_t$ onto $\HH$. Further there exists a continuous process $(\gamma_t)_{t\geq0}$ with values in $\overline{\HH}$ such that
$\HH\setminus K_t$ is the unbounded connected component of
$\HH\setminus\gamma[0,t]$ with probability one. This process is
the trace of the SLE${}_{\kappa}$ and it can be recovered from
$g_t$, and therefore from $W_t$, by
\begin{equation}
\gamma_t  =  \lim_{z\rightarrow W_t, z\in\HH} g_t^{-1}(z)~.
\end{equation}
The constant $\kappa$ characterises the nature of the resulting curves.  For $0<\kappa\leq 4$, SLE${}_{\kappa}$ traces over simple curves, for $4<\kappa<8$ self-touching curves (curves with double points, but without crossing its past) and, finally, if $8\leq\kappa$ the trace becomes space filling. 

Now, for another simply connected domain $D$ with two boundary points $A,B\in\partial D$ the chordal $SLE_{\kappa}$ in $D$ from $A$ to $B$ is defined as
\begin{displaymath}
K_t(D_{A,B}):=h^{-1}(K_t(\HH,0,\infty))
\end{displaymath}
where $K_t(\HH,0,\infty)$ is the hull as in (\ref{Khull}) and $h$ is the conformal map from $D$ onto $\HH$ with $h(A)=0$ and $h(B)=\infty$.

Before we end this section, let us rewrite~(\ref{lowner}) in It{\^o} form, by setting $f_t(z):=g_t(z)-W_t$, which now satisfies the stochastic differential equation
\begin{equation}
\label{Loewner-Ito}
df_t(z)=\frac{2}{f_t(z)} dt-dW_t~.
\end{equation}
For a non-singular boundary point $x\in\R$, we can read off the generator $A$ for the It{\^o}-diffusion $X_t:=f_t(x)$ as 
$$
A=2\frac{1}{x}\frac{d}{dx}-\frac{\kappa}{2}\frac{d^2}{dx^2}~.
$$
Defining the first order differential operators 
\begin{equation}
\label{Witt}
\ell_n:=-x^{n+1}\frac{d}{dx}\qquad n\in\Z~,
\end{equation}
we obtain 
$$
A=\frac{\kappa}{2}\ell^2_{-2}-2\ell_{-1}~.
$$
Let us note, that the differential operators~(\ref{Witt}) form a representation of the Witt algebra~\cite{FW1, FW2}. We shall come back to this matters latter, where we shall see, how the $A$-harmonic functions correspond to null-vectors in a Verma module of the Virasoro algebra. 
\section{Renormalisation Group flow and conformally invariant measures}
\label{RG_sec}
Before we proceed, we shall recall Zamolodchikov's  $c$-Theorem for two-dimensional field theories~\cite{Z}. 

Let us be given a (Euclidean) Field Theory with action functional $S[{\bf g},a]$ depending on an (infinite) set of dimensionless parameters, ${\bf g}=(g_1, g_2,\dots)$, the ``coupling constants", and  an (ultraviolet) ``cut-off" $a$ such that the action is obtained as an integral of local densities, i.e. $S=\int \sigma({\bf g}, a, x) dx$.

The fundamental assumption is the existence of a one-parameter group of motions $R_t$ in the space $Q$ of coupling constants ${\bf g}$, $R_t:Q\rightarrow Q$, with the property that a field theory described by an action $S[R_t{\bf g}, e^t a]$ is equivalent to the original theory with the action $S[{\bf g}, a]$ modulo correlations. This means that all correlation functions calculated in the two theories are the same at scales $x\gg e^ta$ and $t>0$.
The components of the vector fields which generate the renormalisation group (RG) flow are called ``$\beta$-functions", i.e.
\begin{equation}
\label{Z1}
\frac{dg_i}{dt}=\beta_i({\bf g})~.
\end{equation}

Then the following properties hold true for the RG:
\begin{enumerate}
  \item There exists a positive function $c(g)\geq 0$ which decreases monotonically, i.e., 
  \begin{equation}
\label{Z2}
\frac{d}{dt}c=\beta_i(g)\frac{\partial}{\partial g_i}c(g)\leq 0
\end{equation}
with equality only obtained at the fixed points of the RG-flow, i.e., at $g=g_*$, $(\beta_i(g_*)=0)$.
  \item The ``critical" fixed points are stationary for $c(g)$, i.e. $\beta_i(g)=0\Rightarrow \partial c/\partial g_i=0$. Further, at the critical fixed points the corresponding $2D$ field theory is a Conformal Field Theory, with generators $L_n$, $n\in\Z$, of the infinite symmetry algebra, the Virasoro algebra, satisfying the commutation relations 
  \begin{equation}
\label{Z3}
[L_n, L_m]=(n-m) L_{n+m}+\frac{\tilde{c}}{12}(n^3-n)\,\delta_{n+m, 0}~,
\end{equation}
with $\tilde{c}$ the ``central charge" and which is a function of the fixed points, i.e., $\tilde{c}=\tilde{c}(g_*)$.
 \item The value of $c(g)$ at the fixed point $g_*$ coincides with the corresponding central charge in~(\ref{Z3}), i.e., $c(g_*)=\tilde{c}(g_*)$.
\end{enumerate}
\begin{figure}[htbp!]
\begin{center}
\includegraphics[scale=0.4]{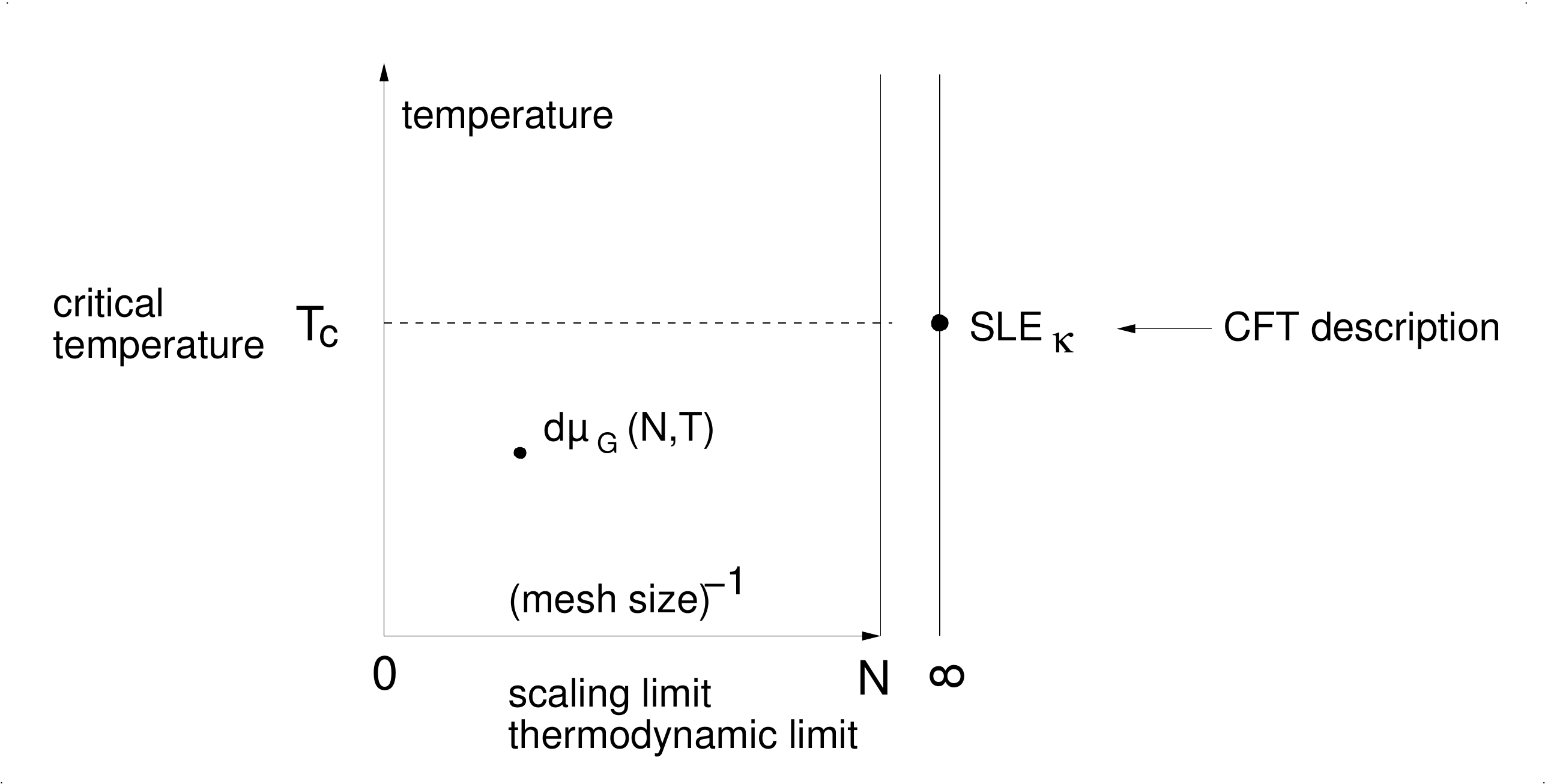}
\caption{The $\mbox{SLE}_{\kappa}$ claim: $\mbox{SLE}_{\kappa}$ is the unstable fixed point of the renormalisation group flow.} 
\label{SLE_RG}
\end{center}
\end{figure}
Now, in Figure~\ref{SLE_RG}, we have summarised the process of approaching the scaling limit of the two-parameter family of Gibbs measures $d\mu_G(N,T)$, $N\sim1/\delta^2$, for a given Boltzmann weight, i.e., for a particular lattice model. Then, as just discussed, the corresponding fixed point in the RG flow is described by a CFT, characterised by its central charge $c$. Additionally the limiting measure should be conformally invariant and supported on random simple chordal paths, i.e., by SLE for some $\kappa$. 

Now, the ``Main Identity" which functionally relates Schramm's diffusion constant $\kappa$ of the SLE process with CFT, has been derived by various people very early~\cite{BB, FW2, K}. The identity reads 
\begin{equation}
\label{kappa=c}
c=\frac{(\kappa-6)(3\kappa-8)}{2\kappa}~.
\end{equation}

Therefore, we know from the $c$-Theorem how to relate the RG flow and SLE. 

The link we have just derived here is of considerable mathematical interest, as it connects dynamical systems, representation theory and stochastic analysis very deeply. It is certainly worth, to be pursued further.

\section{The String model for the Wilson loop and SLE}
The Functional Integral approach to Schramm-L{\oe}wner Evolutions was introduced in ~\cite{F, FK, K} and further extended in~\cite{BF_corr} to Liouville Theory.

The physical problem for the Wilson loop in (Classical) String Theory is to sum over all two-dimensional real surfaces having a closed Jordan curve $C$ in $\R^n$ as boundary.

Historically, this sum $Z$, the partition function, has been defined and calculated (approximately) according to the classical ``Dual String" methods, by applying   the Nambu-Goto action, by Eguchi, Durhuus, Olesen, Nielsen, Petersen, Brink, Di Vecchia, Howe, Deser, Zumino, Lüscher, Symanzik and Weisz for ``dual strings" however it was Polyakov's approach and his specific action, centred around the Dirichlet integral,  which made the problem much more approachable. His action was then subsequently extended to the same problem, i.e, to non-closed strings, by others like Friedan, but decisively by O. Alvarez~\cite{A}. In any case, the underlying physical object is a Functional  Integral, with the action being the specific part, and the mathematical problem left over, to interpret it ``rigorously". 
For mathematical details of String Theory, see, e.g.~\cite{AJPS}.

The ``partition function" is obtained from the well-known functional integral
\begin{equation}
\label{partition_integral}
Z:=\int_{\{g\}}\int_{\{X\}} e^{-S[g, X]}\, [Dg][DX]~,
\end{equation}
which is properly computed over all embeddings $\{X\}$ and all Riemannian metrics $\{g\}$.

The action $S=S[g, X]$, for surfaces $\Sigma$ with boundary $\partial \Sigma$, contains besides the Dirichlet Energy $D[g,X]$ of the embedding, also other terms, to make the theory renormalisable,  and in its simplest version reads in local co-ordinates, 
\begin{eqnarray}
\label{PA_action}
S[g,X] & := & D[g,X]+\text{metric terms} \\
 & = & \frac{1}{2}\int_{\Sigma} g^{ij}\frac{\partial X^{\mu}}{\partial x^i}\frac{\partial X^{\mu}}{\partial y}\sqrt{\det g}\, dx dy+\frac{1}{2\pi}\int_{\Sigma} K_g\sqrt{\det g}\, dx dy+\frac{1}{2\pi}\int_{\partial \Sigma} k_g |dx| 
\end{eqnarray}
where $K_g$ is the Gauss curvature of the metric $g$ and $k_g$ the geodesic curvature of $\partial M$ according to $g$. 

Subsequently we shall not discuss the metric contributions further as they basically enter via the Gauss-Bonnet theorem. Also, we shall only treat regular metrics, although singularities have a profound effect~\cite{BF_corr, KS}. 

Now, in Conformal Field Theory, the marginals, i.e, the integral over all embeddings for a fixed metric is the important quantity, as we recall. This will also reveal the analogy with our earlier discussion in the statistical mechanics approach.
\subsection{Spaces of mappings and the $H^{1/2}$ space on the circle}
The Sobolev space $H^{1/2}(S^1,\R)/\R$
of $L^2(S^1)$ real functions with mean-value zero on the circle can be identified with the sequence space
$$
{\ell}_2^{1/2}=\{~u\equiv(u_0, u_1, u_1,\dots)~|~u_i\in\C~\text{and}~ \{\sqrt{n}\,u_n\}~\text{is square summable}~\}~.
$$
Part of the importance of this Hilbert space comes from the facts that one can interpret its vectors as boundary values of real harmonic functions on the unit disc, $\D$, with finite Dirichlet energy but also as it characterises the subset of quasi-symmetric (qs.) homeomorphisms of the set of all homeomorphisms of $S^1$,~\cite{NS}. The Poisson integral representation gives then a harmonic extension of the space $H^{1/2}$, which is also an isometric isomorphism of Hilbert spaces. For the disc the extension can explicitly be written down in terms of the formula of Douglas.

Let us consider a compact real two-dimensional smooth surface $\Sigma$ with non-empty and non-degenerate boundary $\partial \Sigma$, homeomorphic to the unit circle. Let us also fix a metric $g$, the background metric, on $\Sigma$ and let us consider a map 
$$
h:\Sigma\rightarrow\R^n~,
$$ 
mapping $\partial \Sigma$ diffeomorphically and orientation preserving onto the contour $C$. 

For $h$ harmonic with respect to the metric $g$ we would have for the Laplacian $\Delta_g$,
\begin{equation}
\label{J2.4.1}
\Delta_{g} h\equiv 0.
\end{equation}
The fact which permits to progress further in the endeavour of defining the path integral is first that any embedding $X:\Sigma\rightarrow\R^n$ compatible with the boundary conditions, i.e., $X(\partial\Sigma)=C$ can be decomposed as
\begin{equation}
\label{J2.4.2}
X=h_X+X_0
\end{equation}
where $h_X$ is the unique harmonic map with $h_X|_{\partial\Sigma}\equiv X|_{\partial\Sigma}$ and $X_0|_{\partial\Sigma}=0$. This yields the 
affine space 
$$
h_X+H^{1/2}_0(\Sigma, \R^n)
$$
with $H^{1/2}_0(\Sigma, \R^n)$ denoting the Sobolev space of all maps from $\Sigma$ to $\R^n$ with vanishing boundary values.

An appropriate Hilbert space of states would be  ${\cal H}:=H^{1/2}(\partial\Sigma, C)$.
\begin{figure}[htbp!]
\begin{center}
\includegraphics[scale=0.4]{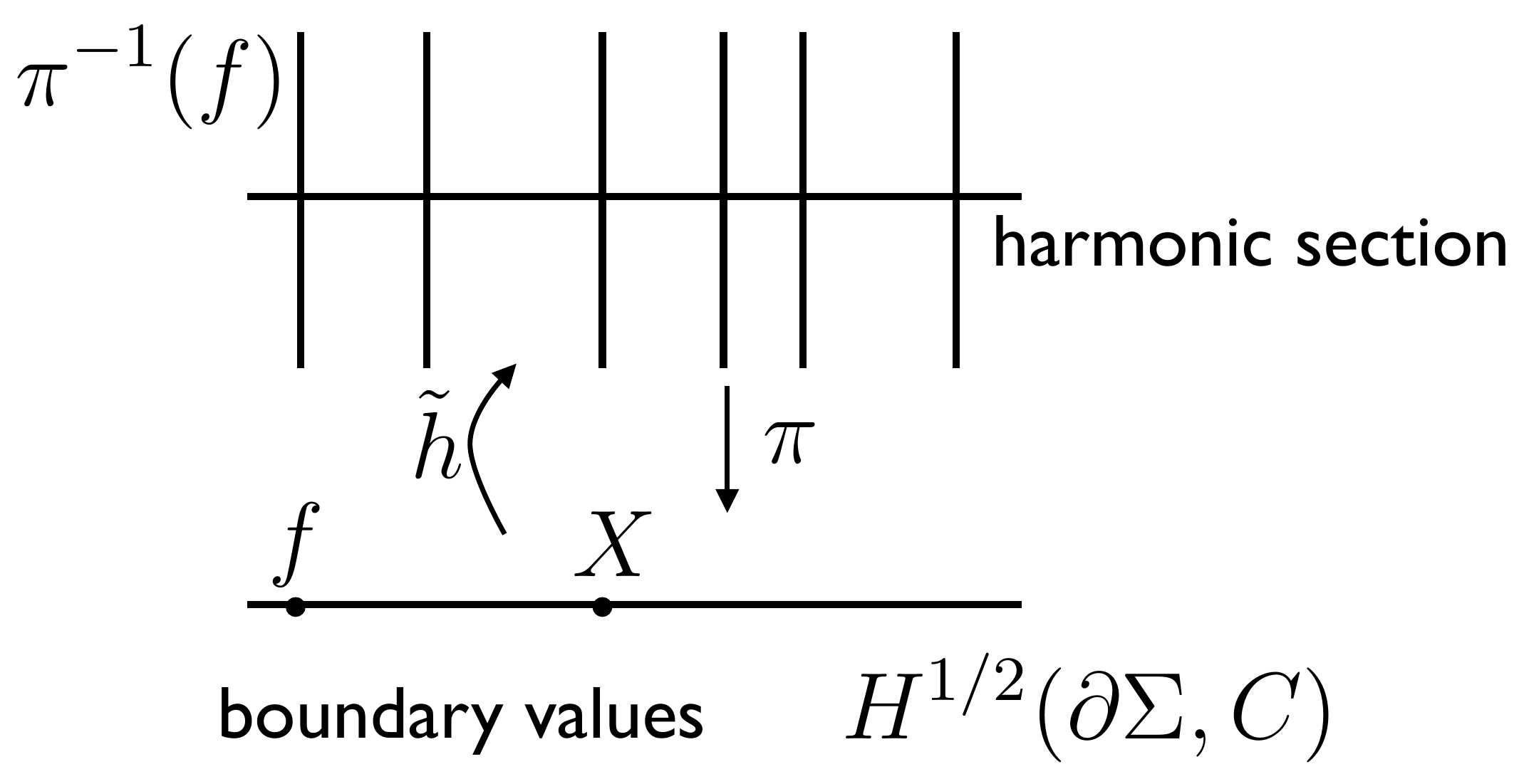}
\caption{The harmonic section $\tilde{h}$ (harmonic extension) of the space of all embeddings of the boundary $\partial\Sigma$ of the surface. The boundary values compose the space of states.} 
\label{h-section}
\end{center}
\end{figure}
Let $f\in {\cal H}$ and let us denote by $\tilde{f}$ its harmonic extension, i.e. the unique harmonic function with boundary value $f$. The situation is schematised in Figure~\ref{h-section}.

In the next few lines we shall deal only with one component of the field. The integral over the fibre $\pi^{-1}(f)$, (cf. Fig.~\ref{h-section}),
\begin{equation}
\label{ }
\Psi[f]:=\int_{\pi^{-1}(f)} e^{-S[g,X]}\, [DX]
\end{equation}
is the continuum version of the partition sum with a specific choice of boundary conditions.  Then the following calculation, with the previous notational conventions, yields for the Dirichlet energy  ($*$ Hodge star),   
\begin{eqnarray}
2\cdot S[X] & = & \int_{\Sigma} dX\wedge *dX =\int_{\Sigma}d(\tilde{f}+X_0)\wedge * d(\tilde{f}+X_0)\\
 & = & \int_{\Sigma} d\tilde{f}\wedge* d\tilde{f}+\int_{\Sigma} dX_0\wedge * dX_0+2\int_{\Sigma} dX_0\wedge d\tilde{f}
\end{eqnarray}
Since $\Delta_g \tilde{f}=d*d\tilde{f}=0$ by definition for the harmonic function $\tilde{f}$, we obtain 
\begin{equation}
\label{ }
S[\tilde{f}+X_0]=S[X_0]+S[\tilde{f}]=S[X_0]+\frac{1}{2}\int_{\partial \Sigma}f*d\tilde{f}~.
\end{equation}
Therefore the partition function factorises into
$$
\int_{\pi^{-1}(f)} e^{-S[g,X]}\, [DX]= e^{-\frac{1}{2}\int_{\partial \Sigma}f*d\tilde{f}}\cdot \int_{\{X_0\}} e^{-S[X_0]}\, [DX_0]~,
$$
which gives (for one component)
\begin{equation}
\label{SurfaceState}
\Psi[f]=\left[\frac{\det(\Delta_g)}{\operatorname{Area}(\Sigma,g)}\right]^{-1/2} \,\cdot\,\e^{-\frac{1}{2}\int_{\partial \Sigma} f* d\tilde{f}}~,
\end{equation}
and where it is understood that the determinants are regularised.  Note that we are dealing with bordered surfaces, and therefore we do not have zero modes.  
 
To obtain the full marginal $Z[g]$, we have to integrate over all possible boundary values. The appropriate measure $\mu$, has been considered   from a Gaussian point of view by G.~Segal and I.~Frenkel, and for ``Unitarising Measures for the Virasoro algebra", by P.~Malliavin, H. Airault and A. Thalmaier~\cite{M, AMT}; (cf. the support of measures in~\cite{AMT}). 
Technically
\begin{equation}
\label{SFM}
\int_{\tilde{h}({\cal H})} e^{-D[g,\tilde{h}]}\,[D\tilde{h}]:=\int_{\varphi\in\operatorname{Hom}_{qs}(S^1)} e^{-\operatorname{E}[\tilde{\varphi},g_0]} d\mu_{\text{SFM}}(\varphi)=:j[{\bf t},C]
\end{equation}
where we have denoted the measure on $\text{Hom}_{qs}(S^1)$ by $d\mu_{\text{SFM}}$, $E[\tilde{\varphi},g_0]$ stands for the Dirichlet integral of the harmonic extension induced by $\varphi$, which also requires now $\Sigma$ to have a complex structure, and finally ${\bf t}$ denotes the conformal class of the metric $g_0$ compatible with the structure on $\Sigma$, i.e. a point in Teichmüller space (cf.~\cite{AJPS}). 

The expression~(\ref{SFM}) is independent of the choices made, with the crucial exception that it depends on the conformal class of the metric.

So, for every fixed metric $g$, the marginal integral in~(\ref{partition_integral}) over embeddings equals 
\begin{equation}
\label{Det_zeta}
Z[g]=\int_{\operatorname{Emb}(\Sigma)} e^{-S[X,g]} [DX]:=j[{\bf t},C]\cdot\left[\frac{\det(\Delta_g)}{\operatorname{Area}(\Sigma,g)}\right]^{-n/2}
\end{equation}

If we restrict to ``planar" surfaces with the Wilson loop $C$ as boundary, i.e. to embeddings of the unit disc into $\R^n$, then the dependence on the conformal class is simple, as all complex structures on the disc are equivalent. It gives also (part) of the relevant partition function for SLE.

However, the regularisation procedure introduces the so-called conformal anomaly, which renders the expressions covariant. Also, for singular metrics interesting effects show up which correspond to exotic versions of the SLE-process~\cite{BF_corr}, e.g. SLE{}$(\kappa,\rho)$~\cite{LSW},  or measure the degree of non-commutativity of SLE, cf.~\cite{FK, LSW}.  

Now, in CFT the important objects are the marginals $Z[g]$, and they are naturally grouped together as a determinant line bundle, as we shall discuss next. 

\subsection{Determinant line bundles and flat connections}
In String Theory but also in Conformal Field Theory, the partition function is considered to be a section of a determinant line bundle. Here we shall briefly recall how one can derive a measure on random paths, by using regularised determinants~\cite{FK, F, K, KS}.

Since to every Jordan domain we can associate the determinant of the Laplacian (with respect to the Euclidean metric and Dirichlet boundary conditions), i.e.,  
$\det(\Delta_D):=\det(\Delta_{g_{\operatorname{Eucl.}}})$,   
we get a trivial bundle over ${\cal M}$, where $f\in{\cal M}$ denotes the uniformising map from the unit disc $\D$ onto the domain $D$, containing the origin.
\[
\begin{CD}
\det(\Delta_{f(\D)})\\
@V \pi VV  \\
 {\cal M} 
\end{CD}
\]
The uniformising map provides us also with a natural connection which allows us to compare the regularised determinants at different points. It has its origin in Polyakov's string theory~\cite{P} and was then subsequently extended in the works~\cite{A,OPS,HZ}.   
Let us consider the space ${\cal F}$ of all flat metrics on $\D$ which are conformal to the Euclidean metric, obtained by pull-back. Namely, for $D\in\text{\texttt{JDom}}$ let $f:\D\rightarrow D$ be a conformal equivalence, and define
$$
\phi:=\log|f'|~.
$$
This gives a correspondence of harmonic functions on $\D$ with the category $\text{\texttt{JDom}}$ and by Weyl rescaling with ${\cal F}$ via 
$$
ds=|f'||dz|=e^{\phi}|dz|~.
$$
To fix the $\operatorname{SU}(1,1)$-freedom, which gives classes of isometric metrics, we divide the state space $H^{1/2}(S^1)$ by the Möbius group of the disc. Henceforth we shall work with the equivalence classes, so, e.g. $0$ corresponds to the orbit of the Euclidean metric under $\operatorname{SU}(1,1)$.

The connection reads~\cite{OPS}:
\begin{equation}
\label{PA_rel}
\det(\Delta_D)=e^{-\frac{1}{6\pi}\oint_{S^1}(\frac{1}{2}\phi{*}d\phi
+\phi|dz|)}\cdot\det(\Delta_{\D})
\end{equation}
Next we would like to show a group property which is essential, as it translates latter into the Markov property, on which conformally invariant measures on paths hinge.  Let us consider the sequence of conformal maps between domains $\D, D, G$:
\[
\begin{CD}
\D@>f>> D@>g>>G~.
\end{CD}
\]
Then the relation of $\det(\Delta_G)$ and $\det(\Delta_\D)$ is obtained via 
$
\frac{d}{dz}g(f(z))=g'(f(z))\cdot f'(z),
$
and
$$
\log|g'(f(z))\cdot f'(z)|=\underbrace{\log|g'(f(z))|}_{=:\psi(z)}+\underbrace{\log|f'(z)|}_{=:\phi(z)}~.
$$
Further by using the property of harmonic functions, i.e., 
$$
\oint_{S^1}\frac{1}{2}(\phi\partial_n\psi+\psi\partial_n\phi)=0~,
$$
gives
$$
\det(\Delta_G)=
\underbrace{e^{-\frac{1}{6\pi}\oint_{S^1}(\frac{1}{2}\psi(f(z)){*}d\psi(f(z))
+\psi(f(z))|dz|)}}_{I.}\cdot\underbrace{e^{\frac{1}{6\pi}\oint_{S^1}(\frac{1}{2}\phi{*}d\phi+\phi|dz|)}\cdot\det(\Delta_{\D})}_{II.}
$$
where
\begin{eqnarray*}
I. & = & \oint_{\partial D}(\frac{1}{2}\tilde{\psi}{*}d\tilde{\psi}~
+\tilde{\psi}|dw|)\qquad\text{with~ $\tilde{\psi}(w):=\log|g'(w)|$}~, \\
II. & = & \det(\Delta_D) 
\end{eqnarray*}
This also shows, that we can consider the determinant for the unit disc as the origin in an infinite affine space.

Let us note, that one should be careful with some of the signs in the literature (cf.~\cite{A,OPS, HZ}). 
Then according to expression~(\ref{Det_zeta}), for the partition function with two-dimensional target space, we have to take the inverse on both sides in~(\ref{PA_rel}), i.e., 
\begin{equation}
\label{PFunction_rel}
Z_D=e^{\frac{1}{6\pi}\oint_{S^1}(\frac{1}{2}\phi{*}d\phi
+\phi|dz|)}\cdot Z_{\D}~.
\end{equation}
The mapping $f$ which maps the unit disc $\mathbb{D}$ onto the slit unit disc $\mathbb{D}\setminus{[1-\sqrt{t},1]}$, can be given by:
$$
f(z)=\frac{\sqrt{1+t}-\sqrt{t+\left(\frac{z-1}{z+1}\right)^2}}{\sqrt{1+t}+\sqrt{t+\left(\frac{z-1}{z+1}\right)^2}}~.
$$
But for such singular disturbances of the boundary the variation formula~(\ref{PA_rel}) ``breaks down", as the Weyl rescaled metric acquires singularities on the boundary.  In order to compute nevertheless the variation of the partition function,  one has to resort to a regularisation procedure, which involves the Schwarzian derivative (cf.~\cite{FK}).

However, as the expression~(\ref{SurfaceState}) showed, we have also to specify boundary conditions, i.e. a state in which our system should be. 

So, if we translate the ``domain-wall" boundary conditions, we get another contribution, which contains the information of the points where the boundary conditions change. In the continuum this corresponds to jump discontinuities along the boundary for the Dirichlet problem. We shall treat here the unit disc with jumps of width $2\lambda$ at $1$ and $-1$. The general case is then obtained by composition with a conformal map. 

The harmonic function compatible with these boundary conditons is
\begin{equation}
\label{harmonic_jump}
u(r e^{it})=\Re\left[\frac{2\lambda}{\pi i}\left(\log(r e^{it}-1)-\log(r e^{it}+1)\right)\right]
\end{equation}
where we have assumed counter-clockwise orientation of the circle. However, the resulting contribution according~(\ref{SurfaceState}) has to be regularised in order to have a finite quantity, $Z_{\lambda}$, depending on the points of discontinuity and the height of the jump(s). But, this also breaks the conformal invariance, such that the quantity varies covariantly over the space of domains, i.e., $Z_{\lambda}=Z_{\lambda}(\text{domain})$ .

\begin{figure}[htbp!]
\begin{center}
\includegraphics[scale=0.4]{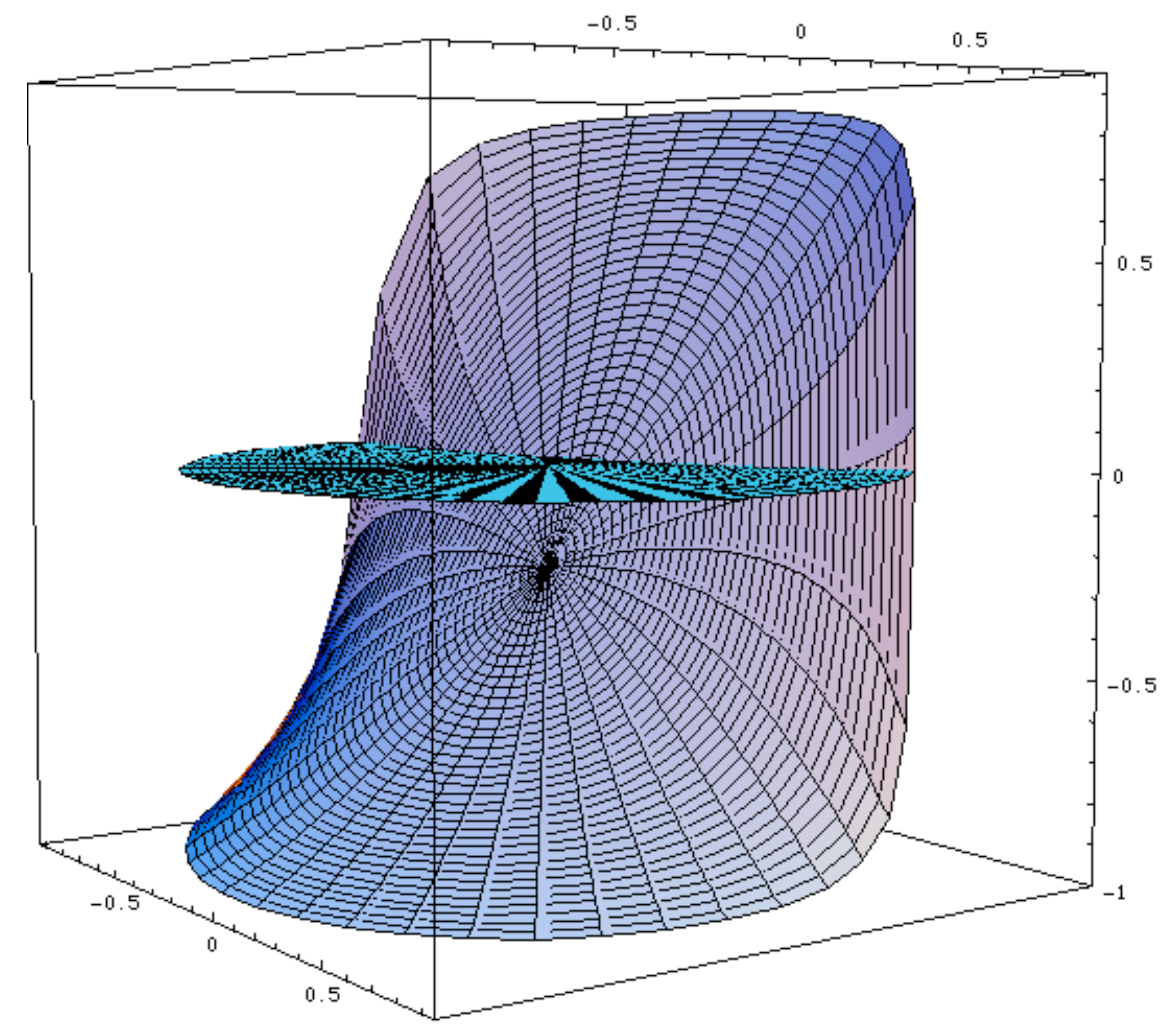}
\caption{Harmonic function which satisfies the jump boundary conditions with $\lambda=1$ on the arc from $\pi/4$ to $\pi$ and $-1$ on the rest, for the Dirichlet problem. In the centre the unit disc is displayed.} 
\label{SLE_RG}
\end{center}
\end{figure}

Nevertheless, this reveals the global geometry of our partition functions in case of the Bosonic free field, i.e. for $c=1$:
\[
\begin{CD}
\label{correlators}
\pi^*(\det(\Delta_D)^{-1})\otimes Z_{\lambda}@>>> {\cal M}\times (S^1\times S^1\setminus\{\text{diagonal}\})\\
@.  @V\pi VV\\
\det(\Delta_D)^{-1}@>>>{\cal M}\hookrightarrow{\aut}({\OO})
\end{CD}
\]

However, what we see is that in the continuum compared with the lattice, there are infinitely many possibilities from which we have to choose the proper ones. The possible choices compose the spectrum and are dictated by the scaling dimensions $h$ of the ``boundary fields", which  can be obtained from highest-weight representations of the Virasoro algebra~\cite{C, BPZ}. They relate to $\lambda$ as $h=\text{const.}\cdot \lambda^2$. In the next section we shall derive this purely from probability theory.  

To summarise, there are a number of variants of SLE, consisting of a Riemannian bordered surface $X$ (oriented, otherwise general topology) with marked points $x_1,\dots, x_n$ on the boundary and $y_1,\dots, y_m$ in the bulk. Analytic coordinates (or merely $1$-jets, i.e. dependence only on the first derivative) at the marked points are given. Then the partition function $Z$ is a positive function of such configurations. It has a tensor dependence on analytic co-ordinates  (i.e. it transforms as $\prod_i(dz_i)^{h_i}\prod_j |dw_j|^{2h_j}$,
$z_i$ local co-ordinate at $x_i$, $w_j$ local coordinate at $y_j$ ), and depends on the metric as $\det(\Delta_D)^{-c}$, where $c$ is the central charge, a real constant. 
Further, it should be positive and equal to the renormalised partition function in the lattice approximation. 

These partition functions are null vectors of canonical Virasoro representations, and  they correspond to correlators in Conformal Field Theory~\cite{FW1, FW2, F, FK, K, KS, BB}. 
The (simplest) version for the disc with two marked points is now stated.

\begin{thm}[2003~\cite{FK, K, F, KS}] Let a configuration $(D, A, B)$, consisting of a simply connected domain $D$, with metric $g$ smooth up to the boundary and two marked boundary points $A, B$, with analytic local co-ordinates, be given. The partition function $Z_{\operatorname{SLE}_{\kappa}}$
of chordal $\operatorname{SLE}_{\kappa}$ is: 
$$
Z_{\operatorname{SLE}_{\kappa}} =  |{\det}_D|^{\otimes c}\otimes |T^*_A\partial D|^{\otimes h}\otimes |T^*_B\partial D|^{\otimes h}=\det(\Delta_D)^{-c}\cdot Z_h\equiv\langle\psi(A)\psi(B)\rangle~, 
$$
where
\begin{eqnarray}
\label{c-charge}
\nonumber
c &=& 1-\frac{3}{2}\cdot\frac{(\kappa-4)^2}{\kappa}=-\frac{(\kappa-6)(3\kappa-8)}{2\kappa}~\text{is the central charge, and}\\
h &=& h(\kappa) =\frac{6-\kappa}{2\kappa}~\text{the highest-weight.}
\end{eqnarray}
$\langle~\rangle$ denotes the unnormalised correlator of the boundary condition changing operators, corresponding to the boundary field $\psi$, of weight $h$.
\end{thm}
Let us comment on the above Theorem. The fundamental and central relation is~(\ref{c-charge}), which ensures that the partition function really behaves as a section of a bundle. Note also the different meanings of $\det$.

Let us discuss the case $\kappa=3$, briefly. According to formula~(\ref{c-charge}) we get $c=1/2$ and  $h=h_{1;2}=1/2$ which would correspond to the Ising model. The weight of the tensor is a half-order differential and the two-point function can be expresses by means of the 
Szeg{\H{o}} kernel for spin $\frac{1}{2}$, i.e., for fermions.  
This little, but important, example should be seen as a  general paradigm which tells us that a correlator with multiple insertions can be derived and expressed in terms of the Green's function, as a Vandermonde determinant, as follows from Wick's theorem.

\section{Doob's $h$-transform, Virasoro null-vectors and SLE}
This is a key section as it gives a unified treatment of the directions initiated by Schramm, Lawler and Werner, and it further connects with the one started by Malliavin. 
Our theory relies on the ``Virasoro Uniformisation" (VU), which in case of the disc, corresponds to the theory developed by A.A. Kirillov, D. Yur'ev~\cite{KY} and Y. Neretin. In addition they took also the underlying infinite Kähler geometry into account.  We shall now build on~\cite{F}. 

\begin{figure}[htbp!]
\begin{center}
\includegraphics[scale=0.4]{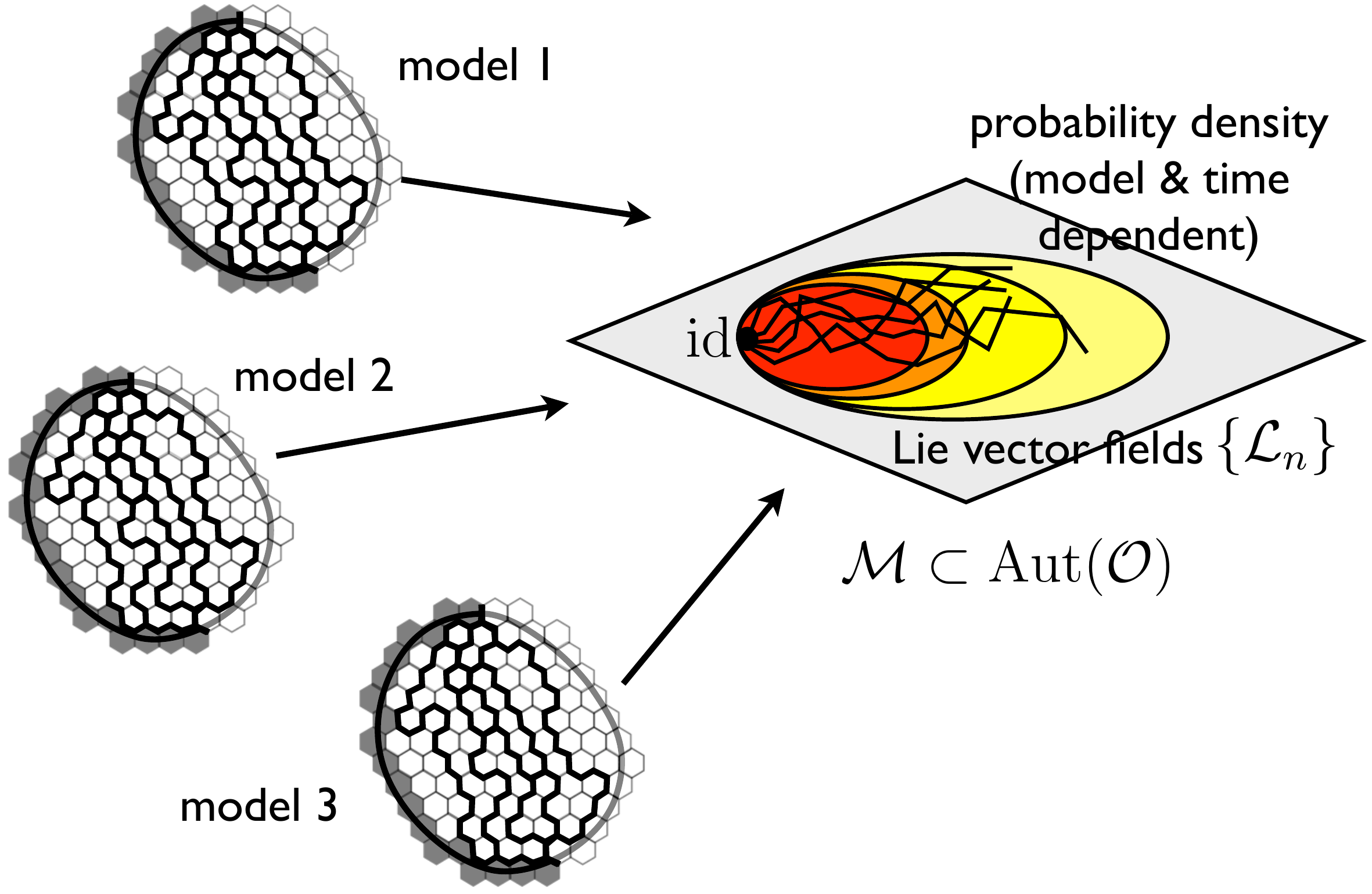}
\caption{Different models with different Gibbs measures produce different probability densities on the same path space in ${\cal M}\subset\aut({\OO})$. The time evolution of the $\kappa$-densities  is governed by a second-order differential operator in Hörmander form, given in terms of the explicit Lie fields ${\cal L}_n$.} 
\label{RW}
\end{center}
\end{figure}

Let us begin by briefly explaining the main idea.  
In Figure~\ref{RW}, we have illustrated the path space which is identical for all models from statistical mechanics, producing simple curves, e.g. domain walls. However, every such model contributes its own measure on the path space. 

If we parametrise it, we get from the measure in the statistical mechanics model different families of probability densities, whose evolution we have again indicated. As our earlier discussion showed, it is the central charge, resp. $\kappa$ which labels different measures. 

Also, to every point corresponds a partition function, which is composed of the regularised determinant and the information on the boundary conditions. The measures on the path space and the transformation properties of this partition functions under the stochastic evolution, i.e., the lifted process, have to form a conformally invariant pair. In particular, the conformal image of any trajectory has to be again a martingale, up to reparametrisation. This requirement gives us the necessary link between the parameters involved, i.e., $\kappa, c, h$, but also explains the appearance of degenerate highest-weight representations of the Virasoro algebra. 

\subsubsection{Path space and Weil-Petersson metric} 
Recall the space ${\cal M}$. An element of it can be written as $c_1z+c_2 z^2+c_3 z^3+\cdots,$ with $c_1>0$. Now, if we fix a parametrisation $t$ of the traces  in our domain, we can cut them accordingly, and obtain so a family of univalent functions of the form
$$
f(t,z)=e^{\gamma(t)}z+c_2(t)z^2+\cdots~,
$$ 
with $\gamma(t)$ a strictly monotone real function reflecting the path along which we are cutting. Further, it follows from the monotonicity that every simple path gives again a simple path in ${\cal M}$. Also, any two different traces in a domain give two different traces in ${\cal M}$, i.e., the map is injective, and unique up to reparametrisation.  This can be seen as follows. 

The traces are compact subsets of the domain, which is a Hausdorff space. If two traces, with the exception of the endpoints are different, there are at least two interior points which are different. Hence, we can choose open disjoint neighbourhoods.  The family of mappings for one of the curves at some instance maps the point onto the boundary, for the other never. Then, it follows from the Open Mapping theorem and the Identity theorem for analytic functions that the corresponding traces in $\aut_+({\OO})$ must have points which are not the same, which in turns shows injectivity.

So, we have that ${\cal M}$ is a subset of the semi-direct product $\R_+\ltimes \Aut_+({\OO})$, where
$$
\Aut_+({\OO}):=z\left(1+\sum_{k=1}^{\infty} c_k z^k\right)~,
$$
and it is enough to study the traces in $\Aut_+({\OO})$. We shall think of the spaces as completed with respect to the natural filtration. The respective Lie groups and algebras are given below:
\begin{eqnarray*}
\Aut_+({\OO}) &  &\qquad\der_+({\OO})=z^2\C[[z]]\partial_z \\
\cap\qquad & & \qquad\qquad\cap \\
\Aut({\OO}) & &\qquad\der_0({\OO})=z\C[[z]]\partial_z \\
& & \qquad\qquad\cap \\
& & \qquad\der({\OO})=\C[[z]]\partial_z
\end{eqnarray*}
and the exponential map $\exp:\der_+({\OO})\rightarrow\Aut_+({\OO})$, being an isomorphism. Further, the group $\Aut({\OO})$ acts on itself by composition.

Now, this space has a natural affine structure with co-ordinates $\{c_k\}$, and the identity map corresponding to the origin $0$.

Let us introduce the following spaces, which are in a sense dual, as they represent the power series developments around infinity, namely
\begin{eqnarray}
\nonumber
\aut({\cal O_{\infty}}) & := & \{~bz+b_0+\frac{b_1}{z}+\cdots~,\quad b\neq0~\}~. \\
\aut_+({\cal O}_{\infty})& := & \{~z+b_0+\frac{b_1}{z}+\cdots~,~\}~.
\end{eqnarray}

Now, the complex Virasoro algebra $\vir_{\C}$ we are considering is spanned by the polynomial vector fields $e_n=-ie^{in\theta}\frac{d}{d\theta}$, $n\in\Z$, and $\mathfrak{c}$, with commutation relations $[\mathfrak{c},e_n]=0$ and 
$$
[e_m, e_n]=[e_m, e_n]+\omega_{c,h}(e_m,e_n)\cdot\mathfrak{c}~,
$$
with the extended  Gelfand-Fuks cocycle
$$
\omega_{c,h}(v_1,v_2):=\frac{1}{2\pi}\int_0^{2\pi}\left((2h-\frac{c}{12})v_1'(\theta)-\frac{c}{12}v_1'''(\theta)\right)v_2(\theta)\,d\theta~,
$$
and $v_1, v_2$ being complex valued vector fields on $S^1$.
It has been shown~\cite{KY}, that there exists a two-parameter family of Kähler metrics on this space, with the form being at the origin \begin{equation}
\label{Kaehler_m}
w_{c,h}:=\sum_{k=1}^{\infty}\left(2hk+\frac{c}{12}(k^3-k)\right)\,dc_k\wedge d\bar{c}_k~,
\end{equation}
building on the Gelfand-Fuks cocycle. This metrics also  generate the Weil-Petersson (WP) metric. So, we can measure ``the distance of two SLE curves", i.e., their shapes, and more generally, do Differential Geometry on the path space.

\subsubsection{Analytic line bundles $E_{c,h}$ over ${\cal M}$} 

The complex vector fields $e_n$ on the circle, which form the Witt algebra, have a representation in terms of the Lie fields ${\cal L}_{e_n}$ which act transitively on $\aut_+(\OO)$.

However, to have an action of ${\vir_{\C}}$, one has to introduce a determinant line bundle. So, on the infinite complex manifold $\aut_+(\OO)$ there exists an analytic line bundle $E_{c,h}$ which carries a transitive action of the Virasoro algebra. The line bundle $E_{c,h}$ is in fact trivial, with total space $E_{c,h}={\aut_+(\OO)}\times\C$. We can parametrise it by pairs $(f,\lambda)$, where $f$ is a univalent function and $\lambda\in\C$.  

It carries the following action
\begin{equation}
\label{KY2}
L_{v+\tau{\mathfrak c}}(f,\lambda)=({\cal L}_v f, \lambda\cdot\Psi(f,v+\tau{\mathfrak c}))~,
\end{equation}
where 
\begin{equation}
\label{Neretin}
\Psi_{c,h}(f,v+\tau{\mathfrak c}):=h\oint\left[\frac{wf'(w)}{f(w)}\right]^2v(w)\frac{dw}{w}+\frac{c}{12}\oint w^2 S(f,w)\frac{dw}{w}+i\tau c~,
\end{equation}
and where 
$$
S(f,w):=\{f;w\}:=\frac{f'''(w)}{f'(w)}-\frac{3}{2}\left(\frac{f''(w)}{f'(w)}\right)^2~,
$$ denotes the Schwarzian derivative of $f$ with respect to $w$.
The central element $\mathfrak{c}$ acts fibre-wise linearly by multiplication with $ic$. 

We have summarised the geometry in the following commutative diagram
\[
\begin{CD}
\label{trans_action}
0@>>> \C@>>>\vir_{\C}@>>>\operatorname{Witt}@>>>0\\
@VVV  @VVV@VVV@VVV@VVV\\
0 @>>> \C@>>>\Theta_{E_{c,h}}@>>>\Theta_{\cal M}@>>>0\\
@VVV@VVV@VVV@VVV@VVV\\
0@>>> \C@>>>E_{c,h}@>>>{\cal M}@>>>0
\end{CD}
\]
where $\Theta_{X}$ denotes the respective tangent sheaf. Generally, to have an action means to have a morphism from a Lie algebra $\mathfrak{g}$  to the tangent sheaf, and transitive means that the map $\mathfrak{g}\otimes{\OO}_{X}\rightarrow\Theta_{X}$ is surjective. 
Algebraically, the situation corresponds to so-called Harish-Chandra pairs $(\mathfrak{g}, K)$.

\subsubsection{$\vir$-modules and Verma modules $V(c,h)$}
One can endow the space of holomorphic sections $|\sigma\rangle\in{\cal O}(E_{c,h})\equiv\Gamma({\aut_+(\OO)}, E_{c,h})$ of the line bundle $E_{c,h}$ with a $\vir_{\C}$-module structure. 
  
Namely, let ${\cal P}$ be the set of (co-ordinate dependent) polynomials on $\cal M$, defined by 
$$
P(c_1,\dots, c_N):\Aut({\OO})/\mathfrak{m}^{N+1}\rightarrow \C~,
$$
with $\mathfrak{m}$ the unique maximal ideal. In the  classical literature on complex variables the above quotient space is called the ``coefficient body". ${\cal P}$ corresponds to the sections ${\cal O}({\cal M})$ of the structure sheaf ${\OO}_{\cal M}$ of ${\cal M}$ and it 
carries an action of the representation of the Witt algebra in terms of the Lie fields ${\cal L}_{n}\equiv{\cal L}_{e_n}$.

In affine co-ordinates $\{c_n\}$, these fields can be written as~\cite{KY}, e.g.
\begin{eqnarray*}
{\cal L}_n & = & \frac{\partial}{\partial c_k}+\sum_{k=1}^{\infty}(k+1) c_k\, \frac{\partial}{\partial c_{n+k}}~\quad n\geq1~.
\end{eqnarray*}

Now, ${\OO}(E_{c,h})$ can be identified with the polynomial sections of the structure sheaf, i.e. with ${\OO}({\cal M})$, by choice of a free generator, and the previous trivialisation.  

The action of $\vir_{\C}$ on sections of $E_{c,h}$ can then be written in co-ordinates, with the notation $\partial_n\equiv\frac{\partial}{\partial c_n}$, according to formulae~(\ref{KY2}, \ref{Neretin}) as~(cf.~\cite{KY}),
\begin{eqnarray}
\label{Vir_coord}
\nonumber
L_n & = & \partial_n+\sum_{k=1}^{\infty}c_k\, \partial_{k+n},\quad n>0 \\\nonumber
L_0 & = & h+\sum_{k=1}^{\infty} k\, c_k\partial_k~,\\ \nonumber
L_{-1} &=& \sum_{k=1}^{\infty}\left((k+2)c_{k+1}-2c_1c_k\right)\partial_k+2hc_1~,\\
L_{-2} &=& \sum_{k=1}^{\infty}\left((k+3)c_{k+2}-(4c_2-c_1^2)c_k-a_k\right)\partial_k+h(4c_2-c_1^2)+\frac{c}{2}(c_2-c_1^2)~,
\end{eqnarray}
where the $a_k$ are the Laurent coefficients of $1/f$, and $c$ the central charge.

The conformal invariance will naturally lead to highest-weight modules.\\  Now a polynomial $P(c_1,\dots, c_N)\in{\OO}(E_{c,h})$ is a singular vector for $\{L_n\}$, $n\geq1$, if
$$
\left(\partial_k+\sum_{k\geq1}(k+1)c_k\partial_{k+n}\right)P(c_1,\dots, c_N)=0~.
$$
Then, the highest-weight vector is the constant polynomial $1$, and satisfies $L_0.1=h\cdot 1$, $Z.1=c\cdot 1$. 

The dual ${\cal O}^*(E_{c,h})$, i.e. the space of linear functionals $\langle\sigma|$ on ${\cal O}(E_{c,h})$, can again be identified with ${\cal O}({\cal M})$ via the pairing,
$$
\langle P,Q\rangle:=P(\partial_1,\dots,\partial_N)Q(c_1,\dots,c_N)|_{c_1=\dots=c_N=0}~,
$$
and the action of $\vir$ can then similarly be given explicitly in co-ordinates as in equations~(\ref{Vir_coord}), with the roles of $c_k$ and $\partial_k$ interchanged.

Analogously, the dual space ${\OO}^*(E_{c,h})$ can be endowed with a $\vir_{\C}$ action. The singular vector for the corresponding (irreducible) Verma module $V_{c,h}$ with respect to $\{L_{-n}\}$, $n\geq1$, is again the constant polynomial $1$.
 
\subsubsection{Hypo-Ellipticity, sub-Riemannian Geometry and Conformal martingales}
After this preparatory explanations, we are in position to make the link with SLE.

The L{\oe}wner equation~(\ref{Loewner-Ito}) can be seen as a family of ordinary complex stochastic differential equations, labelled by the points of the upper half-plane $\mathbb{H}$. But as the individual processes are not independent, in the sense that at every instant $t$, they constitute a random conformal map for the points, where the individual processes are still defined, the interdependence can be translated into a power series expansion around the point which is the least affected in time, namely the point at infinity, or generally, the point at which the trace is aiming. So, we naturally get a random dynamical system on the infinite coefficient body, i.e. a stochastic process $f_{\infty}(t)$ on univalent maps. 

Now, by taking the projective limit we obtain the generator $\hat{A}_{\infty}$ of the flow on $\aut_+({\OO}_{\infty})$, corresponding to the Lœwner equation (6) for some fixed  $\kappa$,~\cite{BB}:
\begin{equation}
\label{L_generator}
\hat{A}_{\infty}=\varprojlim\left(\frac{\kappa}{2}\frac{\partial^2}{\partial b_1^2}+2\sum_{k=2}^N P_k(b_1,\dots, b_N)\frac{\partial}{\partial b_k}\right)~,
\end{equation}
which is driven by one-dimensional standard Brownian motion, and where the polynomials $P_k$ in the drift vector are defined on the coefficient body, with the $N\times N$ diffusion matrix
$$ 
\left(\begin{array}{cccc}1 & 0 & \dots & 0 \\0 & 0 &  &  \\\vdots &  & \ddots  & \vdots \\0 &  &\dots  & 0\end{array}\right)
$$
The generator of the diffusion process can be written in Hörmander form in terms of the tangent vector fields in the affine co-ordinates $\{b_k\}$, applying the notational conventions for operators acting on polynomial sections of the structure sheaf:
$$
L_1^{\infty}:=\frac{\partial}{\partial b_1},\qquad\text{and}\qquad L^{\infty}_2:=-\sum_{k=2}^{\infty}P_k(\underline{b})\frac{\partial}{\partial b_k}~.
$$
But as they satisfy the commutation relations of the Witt algebra, they span the whole tangent space and since we know that this Lie algebra acts transitively, the strong Hörmander condition is satisfied.  Therefore, the resulting flow is hypo-elliptic~\cite{K,F}, and the corresponding geometry sub-Riemannian.

\begin{figure}[htbp!]
\begin{center}
\includegraphics[scale=0.4]{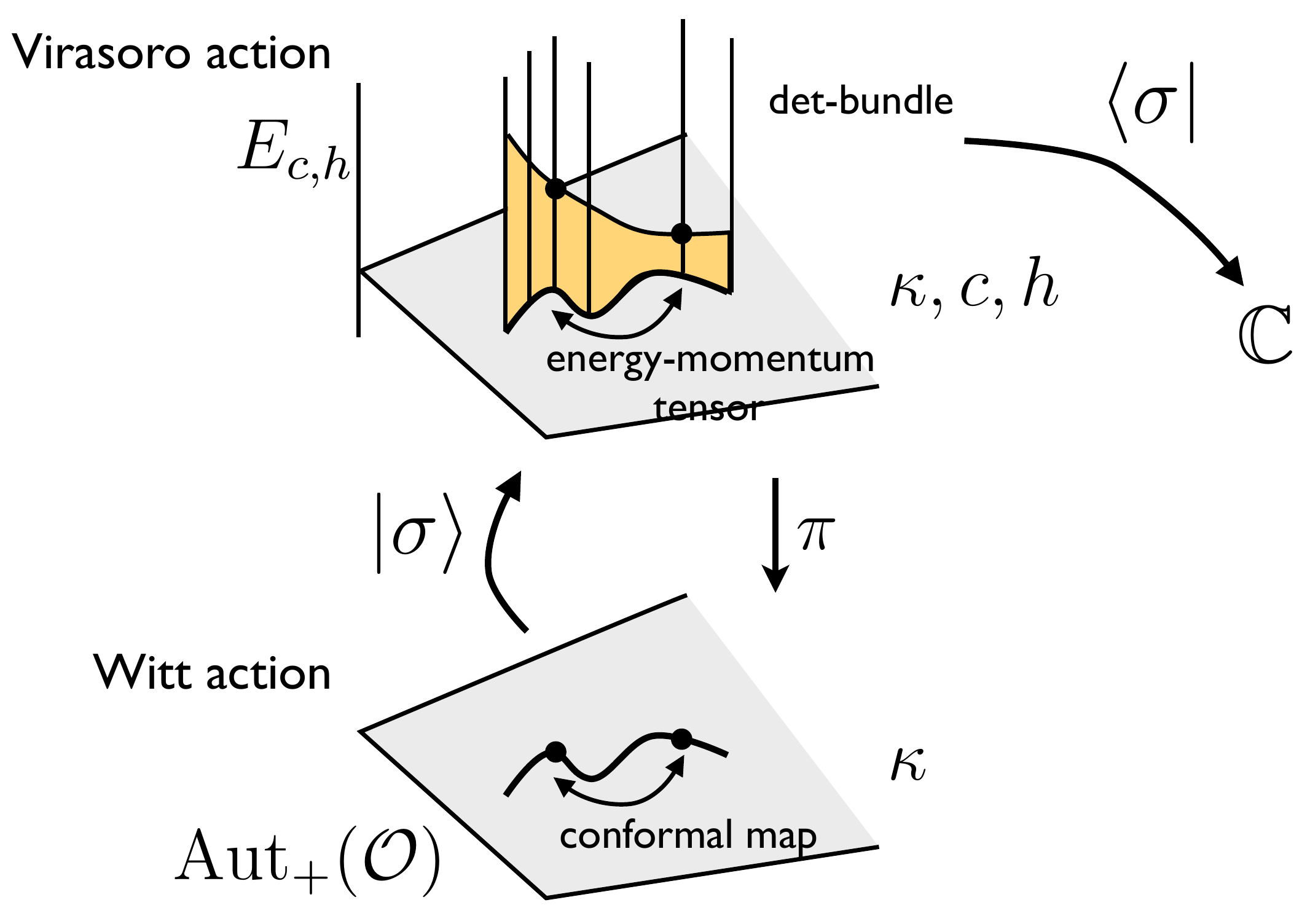}
\caption{The stochastic process in the base $\aut_+({\OO})$ is lifted to the total space $E_{c,h}$ via the section $|\sigma\rangle$. The operators acting on the total space,  depend on the parameters $\kappa, c, h$, coming from the partition function $Z_{c,h}$. The compatibility conditions, i.e., to be a martingale, require $c,h$ to depend functionally on $\kappa$. $\langle\sigma|$ is the dual vector.}
\label{lifted}
\end{center}
\end{figure}

Let us lift the process $f_{\infty}(t)$ on ${\cal M}$ induced by SLE, to the complex manifold $E_{c,h}$ by using sections $\sigma$; (cf. Figure~\ref{lifted}). However, there are some constraints this map has to satisfy, which are given by the transformation properties of the partition function under conformal maps. This means, that the sections have to be flat with respect to the Hermitian connection $\nabla_{c,h}$ (cf.~eq.~(\ref{Kaehler_m})), and the physical connection~(\ref{PFunction_rel}), (energy-momentum tensor), 
which determines $c$ and $h$, for a given model. Again, we have indicted this in~Figure~\ref{lifted}, where the group action which relates different points on the trajectory, corresponds to the transition functions of the L{\oe}wner chain, and in the total space to the conformal anomaly, which relates the points in the fibre, i.e., the values of the determinants over the respective domains. This is equivalent to the Markov property, which was foundational in the derivation of the driving function for SLE. 

Still, there is one more thing. The random trajectories $\sigma_t:=\sigma(f_{\infty}(t))$ should be (local) martingales. This is necessary, if we intend to couple the dynamical picture with the static one, i.e.  ``ensemble averages should be equal to time averages".

To do so, we have  to couple the parameters $\kappa, c, h$,  which we shall obtain from the Doob-Getoor $h$-transform. Namely, if we find harmonic sections, $\sigma_{\text{hr.}}$, then $\sigma(f_{\infty}(t))$ is a local martingale, for the lifted, now Virasoro generators $\hat{L}_n$. 

But, if we restrict to polynomial sections, generated by the $L_n$, $n<1$, by acting on the constant polynomial $1$, which then successively extends over the coefficient body $\Aut({\OO})/\mathfrak{m}^{N+1}$, then as a direct calculation shows, using e.g.~(\ref{Vir_coord}), the module contains a null vector, exactly if
$$
c_{\kappa}=\frac{(6-\kappa)(3\kappa-8)}{2\kappa}\qquad\text{and}\qquad h_{\kappa}=\frac{6-\kappa}{2\kappa}~.
$$
But, this is nothing else than the relation in equation~(\ref{c-charge}), derived from CFT. 
Therefore, all this polynomials are in the kernel of the lifted generator, 
$$
\frac{\kappa}{2}\hat{L}^2_{1}-2\hat{L}_{2}~,
$$
which acts as a differential operator.

Therefore, the representation theoretic notion of ``being degenerate at level two", translates in probabilistic language into a generalised Doob $h$-transform.

This is the way, how conformally invariant measures on simple paths and models from Statistical Mechanics couple, as demonstrated for the special case, the disc. But this is also valid in the general situation.

\subsection*{Conclusions}
We hope that our leisurely promenade has convinced the reader of the beauty and richness of the interplay between the various fields from mathematics and physics. 

The connections which emerged with probability theory are certainly one of the more remarkable facts of this decade. 
Also, the insight that determinant line bundles are a very rewarding starting point to look at SLE, has become, as a glance at the now emerging literature reveals, a `sine qua non` in the field.
However, it is amusing, that although fundamental parts of the material given here have been public for quite some time, all the sudden  people start, e.g. to regularise ``their determinants".   

Finally, it should be immediate that the theory which we presented, builds on an enormous quantity of other theories and mathematical tools, which have been established long before SLE, for other purposes.  

Therefore, SLE is, according to our opinion, yet another good example for the restless ``globalisation of mathematics" and its evolution into a highly integrated organism.


\begin{thebibliography}{99}
\bibitem{AMT} H. Airault, P. Malliavin and A. Thalmaier, {\it Canonical Brownian motion on the space of univalent functions and resolution of Beltrami equations by a continuity method along stochastic flows}, J. de Mathématiques Pures et Appliqués, Vol. {\bf 83}, Issue 8, (2004). 
\bibitem{Ai} M. Aizenman, {\it The geometry of critical percolation and conformal invariance}, StatPhys 19 (Xiamen 1995), (1996).
\bibitem{AB} M. Aizenman A. Burchard, {\it  Hölder Regularity and Dimension Bounds for Random Curves}, Duke Math. J. {\bf 99}, 419 (1999).
\bibitem{AJPS} S. Albeverio, J. Jost, S. Paycha and S. Scarlatti, {\it A mathematical introduction to string theory}, London Math. Soc. LNS {\bf 225}, Cambridge Univ. Press, (1997).
\bibitem{A} O. Alvarez, {\it Theory of Strings with Boundaries: Fluctuation, topology and quantum geometry}, Nucl. Phys. B {\bf 216}, (1983).
\bibitem{BB} M. Bauer and D. Bernard, {\it $2D$ growth processes: SLE and Loewner chains}, Phys. Rep. {\bf 432}, 115 (2006).
\bibitem{BF_corr}
R.~Bauer and R.~Friedrich,
{\it The Correlator Toolbox, Metrics and Moduli},  Nucl. Phys. B {\bf 733}, (2006).
\bibitem{BPZ}
A. A. Belavin, A. M. Polyakov and A. B. Zamolodchikov, {\it Infinite conformal symmetry in two-dimensional quantum field theory}, 
Vol. {\bf 241}, Issue 2, (1984).  
\bibitem{C} J. Cardy, {\it Boundary Conditions, Fusion Rules and The Verlinde Formula}, Nucl. Phys. B {\bf 324}, (1989). 
\bibitem{F1} 
R. Friedrich, preprint, (2008).
\bibitem{F} 
R. Friedrich, {\it On Connections of Conformal Field Theory and 
Stochastic L{\oe}wner Evolution}, arXiv, (2004).
\bibitem{FK}
R. Friedrich and J. Kalkkinen,
  {\it On Conformal  Field Theory and Stochastic Loewner Evolution},
   arXiv 2003, Nucl. Phys. B, {\bf 687}, 279-302 (2004).
\bibitem{FW1}
R. Friedrich and W. Werner,
  {\it Conformal fields, restriction properties, degenerate representations and SLE}, C.R. Acad. Sci. Paris, Ser. I {\bf 335} (2002).
\bibitem{FW2}
R. Friedrich and W. Werner, {\it Conformal  restriction, highest-weight representations and SLE}, Comm. Math. Phys., {\bf 243}, (2003).
\bibitem{GC} A. Gamsa and J. Cardy, {\it Schramm-Loewner evolution in the three-state Potts model--a numerical study}, JSTAT, (2007).
\bibitem{HZ} A. Hassell and S. Zelditch, {\it Determinants of Laplacians in Exterior Domains }, IMRN, Int. Math. Res. Notices, No.{\bf 18}, (1999). 
\bibitem{KY} A.A. Kirillov and D. Yur'ev, {\it Representation of the Virasoro algebra by the orbit method}, JGP, Vol. 5, n. {\bf 3}, (1988).
\bibitem{K} M. Kontsevich, {\it Arbeitstagung 2003--CFT, SLE and phase boundaries},  MPIM2003-60a, (2003).
\bibitem{KS} M. Kontsevich and Y. Suhov, {\it On Malliavin measures, SLE, and CFT}, Proceedings of the Steklov Institute of Mathematics, Vol. 258({\bf 1}), (2007).
\bibitem{LPS} R. Langlands, P. Pouliot, and Y. Saint-Aubin, {\it Conformal invariance in two-dimensional percolation}. Bul l. Amer. Math. Soc., {\bf 30}(1), (1994).
\bibitem{LSW} G. Lawler, O. Schramm and W. Werner, {\it Conformal restriction. The chordal case }, J. Amer. Math. Soc. {\bf 16}, (2003).
\bibitem{M} P. Malliavin, {\it La diffusion canonique au-dessus du groupe des difféomorphismes du cercle}, C.R. Acad. Sci. Paris, Vol. {\bf 329}, Issue 4, (1999).  
\bibitem{NS} S. Nag and D. Sullivan, {\it Teichmüller theory and the universal period mapping via quantum calculus and the $H^{1⁄2}$ space on the circle},   Osaka J. Math. {\bf 32} (1995).
\bibitem{OPS} B. Osgood, R. Philips and P. Sarnak, {\it Extremals of Determinants of Laplacians}, and, {\it Compact Isospectral Sets of Surfaces}, J. Funct. Anal., {\bf 80}, (1988). 
\bibitem{P} A. Polyakov, {\it Quantum geometry of bosonic strings}, Phys. Lett. B, {\bf 103} (1981).
\bibitem{RSch} S. Rohde and O. Schramm, {\it Basic properties of SLE}, Ann. Math. {\bf 161}, (2005).
\bibitem{Sch} O. Schramm, Scaling limits of loop-erased random walks and uniform spanning trees, Israel J. Math. {\bf 118}, (2000).
\bibitem{Sm} S. Smirnov, {\it Towards conformal invariance of $2D$ lattice models}, Proc. Int. Congr. Mathematicians, vol. 2, Eur. Math. Soc., pp. 1421-51, (2006).
\bibitem{TUY} A. Tsuchiya, K. Ueno, and Y. Yamada, {\it Conformal Field Theory on universal family of stable curves with gauge symmetries, in Integrable systems in quantum field theory and statistical mechanics}, Adv. Stud. Pure Math. {\bf 19}, Academic Press, Boston, (1989). 
\bibitem{Z} A.B.~Zamolodchikov, {\it ``Irreversibility" of the flux of the renormalization group in a $2D$ field theory}, JETP Lett. {\bf 43}, (1986). 
\end{thebibliography}
\end{document}